\documentclass[12pt]{article}
\usepackage{amssymb, bbm, amsmath, amsfonts, amsthm, amsbsy, mathtools, mathrsfs, tensor}
\usepackage{enumerate}
\usepackage{graphicx}
\graphicspath{ {Images/} }
\usepackage{color}
\usepackage{hyperref}
\usepackage[marginal, multiple]{footmisc}
\usepackage{cleveref}

\usepackage{upgreek}
\usepackage{euscript}
\usepackage{float}
\usepackage{epsfig}
\usepackage{import}

\newlength{\abstractwidth}
\newcommand {\pd} [2] {\frac {\partial #1} {\partial #2}}

\newcommand {\intf} [2] {\int_{#1}^{#2}}

\newcommand{\inp} [1] {\left( #1 \right)}

\newcommand{\insb} [1] {\left[ #1 \right]}

\newcommand {\adj} [1] {#1^{\dagger}}

\newcommand{\be}{\begin{equation}}
\newcommand{\bea}{\begin{eqnarray}}
\newcommand{\eea}{\end{eqnarray}}
\newcommand{\beq}{\begin{equation}}
\newcommand{\ee}{\end{equation}}

\def \real {\text{Re}\,}

\def\R{\mathbb{R}}
\def\Z{\mathbb{Z}}

\def\eq{&=&}
\def\la{\langle}
\def\ra{\rangle}
\def\t{\tau}
\def\bn{\bigskip \noindent}

\numberwithin{equation}{section}

\theoremstyle{definition}

\theoremstyle{remark}

\begin{document}

\begin{titlepage}

\rightline{}
\bigskip
\bigskip\bigskip\bigskip\bigskip
\bigskip

\centerline{\Large \bf { dS JT Gravity and Double-Scaled SYK }}
\bn
	
\bigskip
\begin{center}
\bf  Adel Rahman  \rm
		
\bigskip
Stanford Institute for Theoretical Physics and Department of Physics, \\
Stanford University,
Stanford, CA 94305-4060, USA
		
\end{center}

\bn
	
\begin{abstract}
\quad

This paper pushes forward a conjecture made in \cite{Susskind:2021esx} that a high-temperature double-scaled limit of the SYK model ($\mathrm{DSSYK}_{\infty}$) describes a de Sitter-like space. We identify a specific bulk theory which we conjecture to be dual to $\mathrm{DSSYK}_{\infty}$, namely JT gravity with positive cosmological constant (dS-JT). We focus our attention on a specific solution of dS-JT in which spacetime is a particular bounded submanifold of dS$_2$ and the profile of the dilaton coincides with that of the radial coordinate of a static patch. This solution can be understood as a dimensional reduction of dS$_3$ and was previously studied by \cite{Svesko:2022txo} in a context different than ours. We describe the geometry of this solution in detail and discuss some ways in which the physics of this solution matches known physics of $\mathrm{DSSYK}_{\infty}$. We describe an example of holographic bulk emergence and find a new role for the timescale $t_* \sim \beta_{\mathrm{GH}}\log(S)$ as the timescale governing this emergence. We discuss some constraints on the boundary-to-bulk operator mapping. This paper provides additional background and context for a companion paper \cite{Lenny} by L. Susskind, which will appear simultaneously.

\end{abstract}
	
\end{titlepage}
\newpage 
\tableofcontents
	
\section{Introduction}

\quad \ This paper follows up on a conjecture made in \cite{Susskind:2021esx} and expanded upon in \cite{Susskind:2022dfz,Lin:2022nss,Lenny} that a high-temperature double-scaled limit of the SYK model (which we will call ``$\mathrm{DSSYK}_{\infty}$") describes a de Sitter-like space. We push the conjecture forward by identifying a specific bulk theory which we conjecture to be dual to $\mathrm{DSSYK}_{\infty}$. The theory is described by two-dimensional Jackiw-Teitelboim gravity \cite{Jackiw:1984je,Teitelboim:1983ux} with positive cosmological constant (which we will call ``dS-JT"). We identify a particular solution of interest which we call dS$_2{}'$ and which we conjecture to be dual to the infinite temperature thermofield double state of the double-scaled SYK model. 

We begin, after briefly defining our framework of static patch horizon holography (as discussed/defined in e.g. \cite{Susskind:2021omt,Susskind:2021dfc,Susskind:2021esx}), by reviewing some basic features of $\mathrm{DSSYK}_{\infty}$, namely the existence of a finite ``tomperature" in the infinite temperature limit, the exponential thermal-like decay of one-sided correlation functions, and the presence of hyperfast scrambling behavior. We use these and other facts to motivate dS-JT as a possible gravitational dual within the context of static patch horizon holography. 

We then go on to study the dS-JT theory---and the dS$_2{}'$ solution in particular---in some detail. We begin our analysis by first considering a ``2D parent solution" for dS$_2{}'$ consisting of the full dS$_2$ hyperboloid and a dilaton whose profile coincides with that of the radial coordinate of a static patch. The solution dS$_2{}'$ is then a restriction of this 2D parent solution to a particular bounded submanifold which is obtained by dimensional reduction of dS$_3$ \cite{Svesko:2022txo}. This alternate origin from a ``3D parent solution" (namely dS$_3$) will strongly influence our analysis. The spacelike\footnote{By ``spacelike dilaton" we mean that $\nabla_a\Phi$ is spacelike---or, equivalently, that the surfaces of constant dilaton are timelike---inside the static patch.} nature of the dilaton in our solution is a key feature which distinguishes our study of dS-JT gravity from previous studies such as \cite{Maldacena:2019cbz,Cotler:2019nbi} (but see also \cite{Svesko:2022txo}). 

We go on to discuss some basic features of this solution, including its temperature, entropy, relation to three-dimensional de Sitter space, and symmetry breaking properties. We then discuss the calculation of one-sided two-point functions of propagating bulk matter, which provides an example of holographic bulk emergence. The timescale $t_* \sim \ell\log(S)$---which would have been the scrambling time in an ordinary fast-scrambling theory \cite{Hayden:2007cs,Sekino:2008he,Maldacena:2015waa}---returns to instead play the role of the timescale at which the full static patch geometry emerges from the holographic boundary theory living at the stretched horizon. We discuss some implications for the boundary-to-bulk operator mapping, finding---using an argument of Susskind and Witten \cite{Lenny}---evidence that the fundamental fermions $\chi_i$ of $\mathrm{DSSYK}_{\infty}$ constitute fundamental near-horizon degrees of freedom rather than propagating bulk fields; the later may still exist in the model as collective degrees of freedom formed from these fundamental fermions, or may be possible to put into the model by hand (see the comment at the end of section \ref{U(1)}).

We conclude by discussing possible extensions of our conjecture and analysis to the charged $\mathrm{DSSYK}_{\infty}$ model and to two-sided spacelike observables. We also stress the need for a better understanding of the ``cosmic sector" of the DSSYK model (as defined here and in \cite{Lenny}, with the factor of $q$ in the Hamiltonian \eqref{DSSYKvsSYK}) at finite $\lambda$.

In this paper, we will take ``$\sim$" to mean ``asymptotic up to scaling by an $O(1)$ constant", as is often done in the literature.

\section{Static Patch Horizon Holography}
\label{framework}
\quad \ Our framework for de Sitter holography will be static patch horizon holography as discussed/defined in \cite{Susskind:2021esx,Susskind:2021omt,Susskind:2021dfc}. In particular, it is assumed that the holographic degrees of freedom encoding a de Sitter static patch reside on the stretched horizon a microscopic distance $\equiv \ell_{\mathrm{s}}$ from the mathematical horizon as in \eqref{rsh} and figure \ref{stretch} below. We assume the microscopic distance $\ell_{\mathrm{s}}$ to be polynomial in the Planck length\footnote{By the ``Planck length" here we mean the inverse of the $(2 + 1)$-dimensional Newton constant. The only place in this paper where the precise location of the stretched horizon plays a role is in the derivation of the timescale $t_* \sim \ell\log(S)$ controlling the emergence of the bulk from the one-sided retarded two-point function.} $\ell_{\mathrm{P}}$ so that 
\begin{equation}
	\log\inp{\ell/\ell_{\mathrm{s}}} \ \sim \ \log\inp{\ell/\ell_{\mathrm{P}}}  \ \sim \ \log(S)
\end{equation}
where $S$ is the de Sitter entropy.
\begin{figure}[H]
	\begin{center}
		\scalebox{0.85}{%% Creator: Inkscape 1.3.2 (091e20e, 2023-11-25), www.inkscape.org
%% PDF/EPS/PS + LaTeX output extension by Johan Engelen, 2010
%% Accompanies image file 'stretch.pdf' (pdf, eps, ps)
%%
%% To include the image in your LaTeX document, write
%%   \input{<filename>.pdf_tex}
%%  instead of
%%   \includegraphics{<filename>.pdf}
%% To scale the image, write
%%   \def\svgwidth{<desired width>}
%%   \input{<filename>.pdf_tex}
%%  instead of
%%   \includegraphics[width=<desired width>]{<filename>.pdf}
%%
%% Images with a different path to the parent latex file can
%% be accessed with the `import' package (which may need to be
%% installed) using
%%   \usepackage{import}
%% in the preamble, and then including the image with
%%   \import{<path to file>}{<filename>.pdf_tex}
%% Alternatively, one can specify
%%   \graphicspath{{<path to file>/}}
%% 
%% For more information, please see info/svg-inkscape on CTAN:
%%   http://tug.ctan.org/tex-archive/info/svg-inkscape
%%
\begingroup%
  \makeatletter%
  \providecommand\color[2][]{%
    \errmessage{(Inkscape) Color is used for the text in Inkscape, but the package 'color.sty' is not loaded}%
    \renewcommand\color[2][]{}%
  }%
  \providecommand\transparent[1]{%
    \errmessage{(Inkscape) Transparency is used (non-zero) for the text in Inkscape, but the package 'transparent.sty' is not loaded}%
    \renewcommand\transparent[1]{}%
  }%
  \providecommand\rotatebox[2]{#2}%
  \newcommand*\fsize{\dimexpr\f@size pt\relax}%
  \newcommand*\lineheight[1]{\fontsize{\fsize}{#1\fsize}\selectfont}%
  \ifx\svgwidth\undefined%
    \setlength{\unitlength}{305.40117159bp}%
    \ifx\svgscale\undefined%
      \relax%
    \else%
      \setlength{\unitlength}{\unitlength * \real{\svgscale}}%
    \fi%
  \else%
    \setlength{\unitlength}{\svgwidth}%
  \fi%
  \global\let\svgwidth\undefined%
  \global\let\svgscale\undefined%
  \makeatother%
  \begin{picture}(1,0.9333452)%
    \lineheight{1}%
    \setlength\tabcolsep{0pt}%
    \put(0,0){\includegraphics[width=\unitlength,page=1]{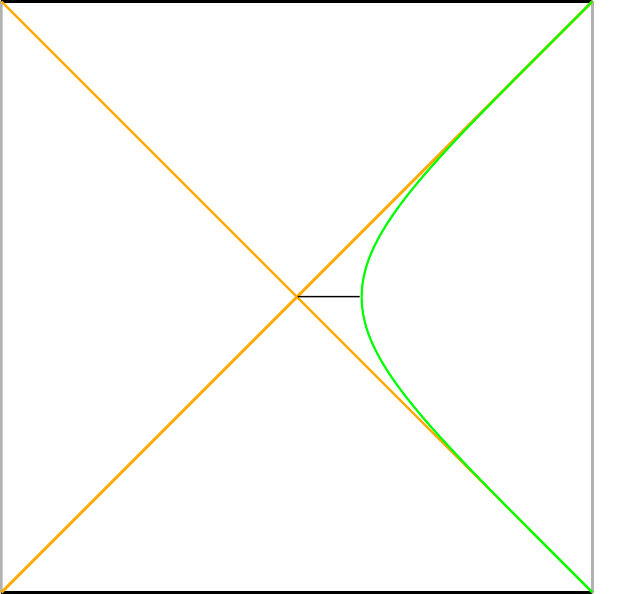}}%
    \put(0.435509,0.67508604){\makebox(0,0)[lt]{\lineheight{1.25}\smash{\begin{tabular}[t]{l}$\delta s \sim O(\ell_{\mathrm{s}})$\end{tabular}}}}%
    \put(0,0){\includegraphics[width=\unitlength,page=2]{stretch.pdf}}%
  \end{picture}%
\endgroup%
}
		\caption{The pode stretched horizon (green) and the mathematical horizon (orange).}
		\label{stretch}
	\end{center}
\end{figure}

Classically, the degrees of freedom describing deformations of the horizon are closely related to quasinormal modes. Examples of quasinormal modes for other types of horizons include viscous flows of the horizon of a black hole and electric currents in the conducting stretched horizon of a black hole. These degrees of freedom are dissipative with a decay rate equal to the imaginary part of the quasinormal mode frequency, which is in turn proportional to the classical bulk temperature. In the quantum theory we assume that there are microscopic degrees of freedom that comprise the entropy \eqref{S} of the horizon/spacetime, out of which the classical degrees of freedom are built. Our main assumption is that when these degrees of freedom are perturbed, the perturbation decays with the quasinormal mode decay rate. In particular we will assume that the time to scramble information is given by the lifetime of the longest lived quasinormal mode (i.e. by the smallest quasinormal mode decay constant), which is generally of order the de Sitter radius \cite{Susskind:2021esx}. 

This is quite different from black hole scrambling where the time scale for scrambling is $\sim R \log(S)$ with $R$ and $S$ the horizon radius and entropy of the black hole respectively. One might have thought that a similar formula would govern scrambling in de Sitter space, but for reasons described in \cite{Susskind:2021esx}  the $\log(S)$ factor is absent. The more rapid scrambling in de Sitter space was described there as ``hyperfast" scrambling. 

Now that we have explained what we mean by static patch horizon holography, we will discuss the boundary and bulk theories which we conjecture to be dual under this framework, namely the high temperature double scaled SYK model ($\mathrm{DSSYK}_{\infty}$) and 2D Jackiw-Teitelboim dilaton-gravity with positive cosmological constant (dS-JT).

\section{Review of DSSYK}
\label{DSSYK}
\quad \ In this section we review some features of the high-temperature limit of the double-scaled SYK model relevant for static patch horizon holography. Much of the discussion follows that of \cite{Susskind:2021esx,Susskind:2022dfz,Lin:2022nss}; those familiar with the content of these references may skip directly ahead to subsection \ref{whydil}. A more detailed discussion of the model in the context of static patch horizon holography is given by \cite{Lenny}.

\subsection{Conventions and Double-Scaled Limit}
\quad \ The holographic boundary theory  we would like to consider is the high-temperature double-scaled limit of the SYK model ($\mathrm{DSSYK}_{\infty}$). There are a few different conventions that are used in the literature for this model; we will follow those of \cite{Lin:2022nss,Lenny}.

The model describes normalized Majorana fermions $\chi_i$ of $N$ different flavors $1 \leq i \leq N$ obeying the Clifford algebra 
\be 
\{\chi_i,\chi_j\} = 2\delta_{ij}
\label{cliff}
\ee
and interacting via the random all-to-all Hamiltonian \cite{Kitaev:2015,Maldacena:2016hyu,Lenny}
\be
H_c[\chi] \ = \ \mathrm{i}^{q/2}\sum_{1\leq i_1 < i_2 \dots < i_q \leq N}j_{i_1i_2\dots i_q}\,\chi_{i_1}\chi_{i_2}\dots\chi_{i_q}
\label{HDSSYK}
\ee 
with real couplings $j_{i_1i_2\dots i_q}$ drawn independently and at random from a Gaussian distribution of mean zero and of variance 
\be 
\la\,j_{i_1i_2\dots i_q}^2\,\ra_{\mathrm{DS}} = \frac{1}{2}\frac{q!}{N^{q-1}}\,\mathcal{J}^2
\label{mcJ}
\ee 
Here $\{\,\cdot\,,\,\cdot\,\}$ denotes the anticomutator and $\mathcal{J}$ is a dimension one parameter which sets the characteristic energy scale of \eqref{HDSSYK}. It will be helpful to think of the Majorana fermions $\chi_i$---and, in particular, the Hamiltonian \eqref{HDSSYK}---as acting on the Hilbert space $\EuScript{H}$ of $\lfloor N \rfloor/2$ qubits.

Our conventions \eqref{HDSSYK}, \eqref{mcJ} differ\footnote{Our conventions also differ from those of \cite{Berkooz:2018jqr}. What we call $\chi$ they call $\psi$, what we call $q$ they call $p$ (they reserve $q$ for a different quantity), what we call $\lambda$ they would call $\lambda/2$, and we have, in the double-scaled limit \eqref{DS} $H_c \sim \sqrt{\frac{2q^2}{\lambda}}\,H^{(\mathrm{Berkooz})} = \sqrt{2N}\,H^{(\mathrm{Berkooz})}$ where we have used our definitions of $q$ and $\lambda$.}
from the usual ones used in e.g. \cite{Maldacena:2016hyu} by two scaling factors. The first, which is purely cosmetic, is\footnote{The usual SYK fermions obey the modified Clifford algebra $\{\psi_i,\psi_j\} = \delta_{ij}$ and hence obey $\psi_i^2 = 1/2$. Our fermions obey \eqref{cliff} and hence obey $\chi_i^2 = 1$.} 
\be 
\chi_i = \sqrt{2}\,\psi_i
\label{chivspsi}
\ee 
where $\psi_i$ are the usual Majorana fermions appearing in e.g. \cite{Maldacena:2016hyu}. The important scaling factor is\footnote{The choices \eqref{chivspsi}, \eqref{DSSYKvsSYK} also imply a rescaling of the form
	\be 
	\la\,j_{i_1i_2\dots i_q}^2\,\ra_{\mathrm{c}} = \frac{q^2}{2^q}\,\la\,J_{i_1i_2\dots i_q}^2\,\ra_{\mathrm{SYK}}^{(\mathrm{usual})}
	\ee}
\be 
H_c[\chi] = q\,H_{\mathrm{SYK}}^{(\mathrm{usual})}[\psi]
\label{DSSYKvsSYK}
\ee 
which is needed for a good double-scaled limit \cite{Lin:2022nss}, at least in the context of describing cosmic scale features (hence the ``c") in de Sitter space \cite{Lenny}, as will be our goal. In particular the Hamiltonians $H_c$ and $H_{\mathrm{SYK}}^{(\mathrm{usual})}$ will give rise to different physics in the large $q$/double-scaled limit where $q \to \infty$. See \cite{Lenny} for further details.

As noted in \cite{Lin:2022nss}, when calculating quantities involving time evolution by $H_c$, the rescaling \eqref{DSSYKvsSYK} is equivalent to keeping the Hamiltonian fixed to its usual form and instead rescaling the time variable conjugate to $H$ via
\be 
\boxed{t_c = \frac{1}{q}\,t^{(\mathrm{usual})}_{\mathrm{SYK}}}
\label{tf}
\ee 
Times which are $O(1)$ in terms of the ``cosmic" clock \eqref{tf} are extremely long---of order $O(q)$---in terms of the usual SYK clock. The advantage of the choice \eqref{tf} is that it allows us to import known formulas for e.g. Green's functions\footnote{Provided these formulas hold for times which are $O(q)$ in terms of the usual SYK clock.} after simply rescaling $t \to qt$.

We will work in the double-scaled limit \cite{Cotler:2016fpe,Berkooz:2018jqr}, which we take to be defined by \cite{Susskind:2021esx,Susskind:2022dfz,Lin:2022nss}
\be 
%q,N \to \infty, \qquad \lambda \equiv \frac{q^2}{N^{p}} = \text{ fixed }, \qquad 0 < p \leq 1
q,N \to \infty, \qquad \lambda \equiv \frac{q^2}{N} = \text{ fixed }
\label{DS}
\ee
%This is a bit more general than the usual definition of the double-scaled limit used in e.g. \cite{Cotler:2016fpe,Berkooz:2018jqr} which only captures the edge case $p = 1$. We will remain agnostic to the precise value of $p$ which, as noted in \cite{Lin:2022nss}, may be further constrained by $1/N$ corrections.
Not much is known about the $H_c$--conjugate dynamics of operators of size $\sim O(q^0)$ in $\mathrm{DSSYK}_{\infty}$. More is known about a related limit of the SYK model which is usually called the ``large $q$" limit \cite{Maldacena:2016hyu}, in which one first takes the large $N$ limit \textit{before} taking $q \to \infty$. There is reason to believe (see \cite{Lenny})\footnote{\cite{Lenny} argues that the parameters $N$, $\lambda$, and $q^2 \equiv N\lambda$ of $\mathrm{DSSYK}_{\infty}$ are closely analogous to the parameters $N^2$, $g^2$, and $\alpha \equiv Ng^2$ ('t Hooft coupling) of $\mathcal{N} = 4$ super Yang-Mills theory (SYM). Working in the ``large $q$" limit $\lambda \ll 1$ in $\mathrm{DSSYK}_{\infty}$ is analogous to working at small gauge coupling $g^2 \ll 1$ in SYM, in which gauge theory calculations are tractable but lead to a nonlocal AdS bulk with string scale parameterically larger than the Planck scale. \cite{Lenny} argues that working in the ``large $q$" limit $\lambda \ll 1$ in $\mathrm{DSSYK}_{\infty}$ similarly leads to a nonlocal de Sitter bulk with a nonlocality scale parameterically larger than the Planck scale; locality emerges as $\lambda \to 1$. However \cite{Lenny} also argues that, as long as $\lambda N \gg 1$, this nonlocality scale will always be parameterically smaller than the cosmic/de Sitter scale $\ell$. This is why we expect the large $q$ and finite $\lambda$ regimes to agree on cosmic/de Sitter scales in their emergent JT de Sitter bulks.} that the large $q$ and double scaled \eqref{DS} limits will agree on cosmic/de Sitter scales in their emergent bulks provided we appropriately scale the Hamiltonian with $q$ as in \eqref{DSSYKvsSYK} in both cases (in fact, the large $q$ limit, as we have defined it here, is simply the edge case $\lambda = 0$ of the double-scaled limit \eqref{DS}). Therefore, in this paper, we will sometimes use the large $q$ and double scaled limits interchangeably, with the understanding that we will only try to recover the resulting data on cosmic/de Sitter scales in holographic bulk, and with the understanding that in our definition of ``large $q$" we are including the rescaling \eqref{DSSYKvsSYK}.

\subsection{High Temperature Limit and Tomperature}
\label{hightemp}
\quad \ In the canonical ensemble, the thermal state at inverse temperature $T^{-1} = \beta$ is described by the density matrix 
\be 
\rho_{\beta} = \frac{1}{Z}\,e^{-\beta H}, \qquad Z \equiv \mathrm{Tr}\inp{e^{-\beta H}}
\label{rhobeta}
\ee
We will often like to think of this as the one-sided reduction of the thermofield double (TFD) pure state\footnote{Here $|\,n\,\ra$ and $E_n$ denote the eigenstates and corresponding energies of $H$ and $^*$ corresponds to the action of an antiunitary transformation $\Theta: \EuScript{H}_R \to \EuScript{H}_L$ which exchanges the two copies of the Hilbert space and, in particular, changes the overall sign of the time variable conjugate to the Hamiltonian
	\be 
	\exp\inp{-\mathrm{i}H_Lt}\,\Theta = \Theta\,\exp\inp{-\mathrm{i}H_R(-t)}
	\label{Theta}
	\ee
	Specifically, we have 
	\be 
	|\,n\,\ra_{L}^* \equiv \Theta\,|\,n\,\ra_{R}^{}
	\label{Thetan}
	\ee} 
\be 
|\,\mathrm{TFD}_{\beta}\,\ra = \frac{1}{\sqrt{Z}}\,\sum_{n}e^{-\frac{\beta}{2}\,E_n}\,|\,n\,\ra_{L}^*|\,n\,\ra_{R}^{}
\label{TFD}
\ee
which lives in the product $\EuScript{H}_{\mathrm{TFD}} = \EuScript{H}_{L}\otimes\EuScript{H}_{R}$ of two copies of the Hilbert space (the ``left" copy $\EuScript{H}_{L}$ and the ``right" copy $\EuScript{H}_{R}$). In the high temperature limit $\beta \to 0$, the TFD becomes a maximally mixed state with flat entanglement spectrum\footnote{The idea that de Sitter space should be described by a quantum-mechanical system at formally infinite temperature/by a state with flat entanglement spectrum has previously been suggested by \cite{Banks:2003cg,Banks:2006rx,Fischler,Dong:2018cuv} and, most recently, \cite{Chandrasekaran:2022cip}.}, and the entropy of the $\mathrm{DSSYK}$ thermal state \eqref{rhobeta} becomes
\be 
S \equiv \frac{\log(2)}{2}\,\lfloor N \rfloor
\label{SDSSYK}
\ee
where $\lfloor\,\cdot\,\rfloor$ is the floor function. In the double-scaled limit \eqref{DS}, \eqref{SDSSYK} scales parameterically with $N$ as 
\be 
\boxed{S \ \sim \ N}
\label{SDSSYKN}
\ee

In thermal states, many time scales---such as the decay rate of two-point functions and the scrambling time---are controlled by the inverse temperature $\beta$. It turns out that for some discrete (fermionic/qubit) systems such as $\mathrm{DSSYK}_{\infty}$, these time scales do not go to zero as $\beta \to 0$ but instead go over to finite nonzero limits controlled by an emergent ``effective temperature" which, following \cite{Lin:2019kpf,Lin:2022nss} we call the ``tomperature" $\mathcal{T}$.

In the microcanonical ensemble, the temperature $T$ is defined by the change in the energy when the entropy is increased by one unit, with all other parameters of the system held fixed. The tomperature $\mathcal{T}$, by contrast, is defined (at least in our context) as the change in the cosmic energy---as defined by the cosmic Hamiltonian \eqref{HDSSYK}---when one qubit (or two fermions) is removed from the system, with the couplings involving all other fermions held fixed. \cite{Lin:2022nss} find that, for $\mathrm{DSSYK}_{\infty}$,
\be 
\boxed{\mathcal{T} =  2\mathcal{J}}
\label{tomperature}
\ee
%See \cite{Lin:2022nss} for the definition of tomperature and for further details.

\subsection{One-Sided Retarded Two-Point Function}
\label{GSYK}
\quad \ Using results from \cite{Maldacena:2016hyu,Roberts:2018mnp,Tarnopolsky:2018env,Maldacena:2018lmt}, one finds that in the high-temperature large $q$ limit, the one-sided retarded two-point function\footnote{Note that the retarded two-point function \eqref{G(t)beta} can be simply related to the more conventional Wightman two-point function $\la\,\chi(t_c)\,\chi(0)\,\ra_{\beta}$ via 
\be 
G_{\beta}^{(\mathrm{R})}(t_c) 
=  2\mathrm{i}\,\Theta(t_c)\,\mathrm{Im}\la\,\chi(t_c)\,\chi(0)\,\ra_{\beta} 
\ee
This simply follows from the fact that (accounting for fermionic statistics)
\be 
\la\,\chi(t_c)\,\chi(0)\,\ra_{\beta}^* = -\la\,\chi(0)\,\chi(t_c)\,\ra_{\beta}
\ee
}
\be 
G_{\beta}^{(\mathrm{R})}(t_c) \equiv  \Theta(t_c)\,\big\la\,\{\chi(t_c),\chi(0)\}\,\big\ra_{\beta}
\label{G(t)beta}
\ee
obeys \cite{Roberts:2018mnp,Maldacena:2018lmt,Lin:2022nss}
\begin{align}
	G_0^{(\mathrm{R})}(t_c) \ \sim \ 
	2\,\Theta(t)\,e^{-2\mathcal{J}\,t_c} + O(q^{-1},\lambda)
	\label{G(t)SYK}
\end{align}
where we emphasize that by first taking the large $N$ limit \textit{before} taking $q \to \infty$, we have automatically localized ourselves to the regime $\lambda = 0$. The correlation function in \eqref{G(t)beta} is a real-time/Lorentzian correlator in the one-sided thermal state \eqref{rhobeta} at temperature $\beta$ and $t$ denotes the clock \eqref{tf} conjugate to the cosmic Hamiltonian \eqref{HDSSYK}. $\Theta(\,\cdot\,)$ denotes the usual Heaviside step function. 

After using \eqref{tomperature}, this simply becomes
\be
\boxed{G_0^{(\mathrm{R})} (t_c) 
	\ \sim \ 2\,\Theta(t_c)\,e^{-\mathcal{T}t_c} + O(q^{-1},\lambda)}
\label{G0(t)}
\ee
In other words, in the large $q$ limit \eqref{DSSYKvsSYK} the one-sided retarded two-point function exponentially decays with a coefficient given by the tomperature. This looks like the IR/late time behavior of an ordinary thermal two-point function, with thermalization time given by the inverse tomperature.

In the context of holography, we expect objects like \eqref{G(t)beta} to encode the ringdown of bulk horizons after small perturbations, as captured by the thermal decay of bulk quasinormal modes \cite{Horowitz:1999jd}. We therefore seek a bulk theory in which finite bulk temperature, due to a warm\footnote{Here by ``warm" we mean that the geometry near the horizon is free to fluctuate at finite $S$.} event horizon, emerges from a boundary theory of infinite temperature but finite tomperature. \cite{Lin:2022nss} showed---in response to the ideas raised in \cite{Banks:2003cg,Fischler,Dong:2018cuv,Chandrasekaran:2022cip}---that de Sitter space in the context of static patch horizon holography is precisely such a candidate bulk. We will explicitly check in section \ref{2ptonesided} below that the leading decaying quasinormal modes of appropriately treated bulk fields in dS-JT encode the same thermal behavior as \eqref{G0(t)}; in the process we will find, using an argument of Susskind and Witten \cite{Lenny}, some evidence suggesting that the fundamental Fermions $\chi_i$ in \eqref{G(t)beta} encode fundamental near-horizon degrees of freedom rather than propagating bulk fields.

\subsection{Hyperfast Scrambling}
\label{hyperfastS}
\quad \ As explained in \cite{Susskind:2021esx,Susskind:2022dfz,Lin:2022nss}, the high temperature $\mathrm{DSSYK}$ model describes a hyperfast scrambler: the growth of operator size with time is governed by a scrambling time which is $O(\mathcal{T}^{-1})$ (in terms of the natural clock variable \eqref{tf}), which is much shorter/faster than the $O(\mathcal{T}^{-1}\log(S))$ expected to bound the scrambling time in a theory with a $k$-local Hamiltonian \cite{Sekino:2008he,Maldacena:2015waa}. The $\mathrm{DSSYK}$ model overcomes the fast-scrambling bound due to its nonlocality: the theory is not $k$-local since the locality parameter $q$ grows with $N$.

To understand hyperfast scrambling, it is helpful to first review operator growth in a standard $k$-local chaotic theory such as SYK with fixed $q$. There are two exponential behaviors of the operator size $\sigma(\tau)$ ($\tau$ the clock variable conjugate to the Hamiltonian of interest). The early time behavior is exponential growth\footnote{To be more precise, this early time behavior is preceded by a transient epoch of length $\delta \tau \lesssim O(q)$. This epoch describes the ``first step" of scrambling in which a single fermion grows into an operator of size $O(q)$. There is a large variance, of order $\delta\tau \sim O(q)$, in how long this first step takes, which serves as a ``bottleneck" on how quickly scrambling actually occurs when $q$ is large in the usual SYK model. In the double-scaled limit \eqref{DS}, and in terms of the cosmic clock $t_{\mathrm{DS}}$, this epoch only lasts for $\delta t_c \lesssim O(1)$ and so does not change the conclusion that scrambling in $\mathrm{DSSYK}_{\infty}$ is hyperfast, taking place on a timescale which is $O(1)$ in the entropy. We thank D. Stanford for explaining this to us.}
\be 
\sigma(\tau) \ \sim \ e^{\lambda \t}
\label{early}
\ee
while the late time behavior is an approach to a plateau
\be
\frac{\sigma(\tau)}{N} \sim 1-e^{-\gamma \t}
\label{late}
\ee
The Lyapunov exponent $\lambda$ is bounded by the chaos bound \cite{Maldacena:2015waa}, while the decay constant $\gamma$ is equal to the smallest quasinormal mode decay constant \cite{Maldacena:2016hyu}.

As detailed in \cite{Susskind:2021esx,Susskind:2022dfz,Lin:2022nss} one can use the continuum limit of the random circuit epidemic model %\footnote{There is some sense in which the continuum limit of the random circuit model may fail to be a good approximation to the underlying discrete-time model. One can do better by instead working with a ``vaccinated" model, in which the qubits only infect each other with probability $\epsilon$ (and, optionally, to make the connection to a continuum limit more manifest, come together after time steps of size $\epsilon$). With $\epsilon < 1$, the continuum limit is a good approximation to the underlying discrete-time model, and one can check numerically that we need not take $\epsilon$ very small in order to achieve good agreement.} 
\cite{Roberts:2018mnp,Susskind:2014jwa} as a proxy for operator growth in $\mathrm{DSSYK}_{\infty}$ to find that in the double-scaled limit, and in terms of the cosmic clock \eqref{tf},
\be 
\frac{\sigma_{\mathrm{DSSYK}}(t_c)}{N} = 1-e^{-\mathcal{T}t_c}
\label{DSscramb}
\ee 
for all positive $t_c > 0$ (here we have also used \eqref{tomperature}). In the high temperature $\mathrm{DSSYK}$ model, the operator growth curve \eqref{DSscramb} is entirely captured by the ``late time" behavior \eqref{late}: the region of Lyapunov behavior \eqref{early}, which persists for about
\be 
\Delta t_{c}^{(\mathrm{Lyapunov})} \ \sim \ \frac{\mathcal{T}^{-1}}{q}\,\log\inp{\frac{N}{q}} \ \underset{\text{double-scaled}}{\to} \ 0
\label{dtly}
\ee
has parameterically shrunk to zero.

The scrambling time ${t_{c}}_*$---i.e. the time at which $\sigma_{\mathrm{DSSYK}}(t_{c}) \sim O(N)$---is easily seen to be 
\be 
{t_{c}}_* \ \sim \ \mathcal{T}^{-1}
\label{t*}
\ee 
which is hyperfast as defined above. By contrast, in the high temperature limit of the ordinary SYK model (with $q$ large but with clock given by $t_{\mathrm{SYK}}^{(\mathrm{usual})}$) we have 
\be 
t_{\mathrm{SYK}*}^{(\mathrm{usual})} \ \sim \ \mathcal{J}^{-1}\log\inp{\frac{N}{q}}
\label{t*usual}
\ee 
consistent with ordinary fast scrambling/$k$-locality.

The upshot is that, unlike ordinary $k$-local theories, $\mathrm{DSSYK}_{\infty}$ exhibits hyperfast scrambling, with operator growth dominated by the dynamics of quasinormal modes (which are in turn controlled by the finite tomperature). In a holographic gravitational dual, we expect operator growth to map onto the dynamics of the duals of simple boundary operators as they fall towards bulk horizons. The behavior \eqref{t*} means that we must seek a bulk theory in which the holographic degrees of freedom live on or near the stretched horizon. de Sitter space, in the context of static patch horizon holography, is precisely such a theory \cite{Susskind:2021esx,Lin:2022nss}. 

\subsection{Why a 2D Dilaton-Gravity Dual?} 
\label{whydil}
\quad \ We arrive at two-dimensional Jackiw-Teitelboim dilaton-gravity with positive cosmological constant as a candidate gravitational dual to $\mathrm{DSSYK}_{\infty}$ by the following (rough and somewhat heuristic) argument\footnote{This argument is originally due to L. Susskind.}: There is nothing to suggest that the $\mathrm{DSSYK}_{\infty}$ model contains anything like tunable-mass Schwarzschild-de Sitter black holes. This suggests a holographic dual of spacetime dimension $D \leq 3$. On the other hand, $\mathrm{DSSYK}_{\infty}$ has a one-parameter tunable entropy $S \sim N$. Horizons in pure 2D Einstein gravity have no entropy since the tranverse area vanishes (the theory also does not have any nontrivial equations of motion). A minimal modification of this theory which would allow for a one-parameter tunable entropy (as well as nontrivial equations of motion) would be to add a dilaton.

In the other direction, three-dimensional de Sitter space in 3D Einstein gravity has many of the features that we see or hope to see in $\mathrm{DSSYK}_{\infty}$ \cite{Susskind:2021esx}, but it seems unlikely that pure $\mathrm{DSSYK}_{\infty}$ would be able to holographically generate both spatial dimensions of dS$_3$ in their entirety (but see also subsection \ref{U(1)}). We can replace one of these spatial dimensions with a single real degree of freedom by dimensionally reducing the 3D theory on the spatial circle. As we will review below, if we began with 3D Einstein gravity with positive cosmological constant, we end up with 2D JT gravity with positive cosmological constant \cite{Svesko:2022txo}.

With this motivation, we will begin to study this conjectured bulk dual.

\section{The Bulk: dS-JT Gravity}
\label{bulk}
\subsection{Action and Solution}
\quad \ The bulk gravitational theory that we would like to consider is described by two-dimensional Jackiw-Teitelboim (JT) dilaton-gravity with positive cosmological constant. The relevant part of the 2D dilaton-gravity action is given by \cite{Jackiw:1984je,Teitelboim:1983ux,Maldacena:2019cbz,Cotler:2019nbi}\footnote{We have (possibly \cite{Svesko:2022txo}) left out the topological term proportional to $\int\sqrt{|g|}R$, which plays no role here.} 
\be
	I_{\mathrm{JT}}[g_{ab},\Phi] = \frac{1}{16\pi G_N^{(2)}}\int_{\mathcal{M}} \mathrm{d}^2x\sqrt{|g|}\,\Phi\inp{R-2/\ell^2}
	+ \frac{1}{8\pi G_N^{(2)}}\int_{\partial\mathcal{M}}\mathrm{d}x\sqrt{|h|}\,\Phi K
	\label{action}
\ee
%Here and below we work in units in which the two-dimensional de Sitter radius $\ell$ is equal to one, unless otherwise stated. 
Here $R$ is the Ricci curvature scalar of the 2D metric $g_{ab}$ and $K$ is the extrinsic curvature scalar\footnote{Calculated with the outward (inward) pointing unit normal for a timelike (spacelike) boundary component. We will only be concerned with timelike boundaries here.} of its restriction, $h$, to the possible boundary $\partial\mathcal{M}$ (on which we impose Dirichlet boundary conditions for the dilaton). The 2D Newton constant $G_N^{(2)}$ is dimensionless. We \textit{define} the 2D Planck length $\ell_{\mathrm{P}}$ via 
\be 
\ell_{\mathrm{P}} = 2\pi\ell\,G_N^{(2)}
\label{ellP}
\ee 
The motivation for doing this will be explained in section \ref{3D} below. For now we simply note that small 2D Newton constant corresponds (linearly) to small 2D Planck length up to a coefficient which is $O(1)$ in de Sitter units.

The dilaton equation of motion, obtained by varying the action \eqref{action} with respect to $\Phi$, sets
\be 
R=2/\ell^2
\label{R=2}
\ee
In Lorentz signature the simplest solution to this equation (though not the one we will ultimately be interested in) is two-dimensional de Sitter space (dS$_2$), which can be represented as a timelike hyperboloid of radius $\ell$ embedded in three-dimensional Minkowski space. If we describe the 3D Minkowski embedding space using global inertial/Cartesian coordinates $X^A = (T, X, Y)$ in the usual way, the de Sitter hyperboloid is described by the equation
\be  
P(X,X) = X^2+Y^2-T^2=\ell^2
\label{hyperboloid}
\ee
where we have defined the de Sitter invariant $P(X,X')$ via 
\be 
P(X,X') \ \equiv \ \eta_{AB}X^AX'^B
\label{PXXp}
\ee 
where $\eta_{AB}$ is the flat metric of the 3D Minkowski embedding space.

More generally, we can represent $D$-dimensional de Sitter space as a hyperboloid in $(D + 1)$-dimensional Minkowski space. Describing de Sitter space in this way allows us to determine the isometry group to be
\be
\mathrm{Iso}(\mathrm{dS}_D) = O(D,1)
\label{dSIso}
\ee
which is simply the subgroup of the embedding space Poincar\'e group which leaves the de Sitter hyperboloid invariant.

\begin{figure}[H]
	\begin{center}
		\includegraphics[scale=.325]{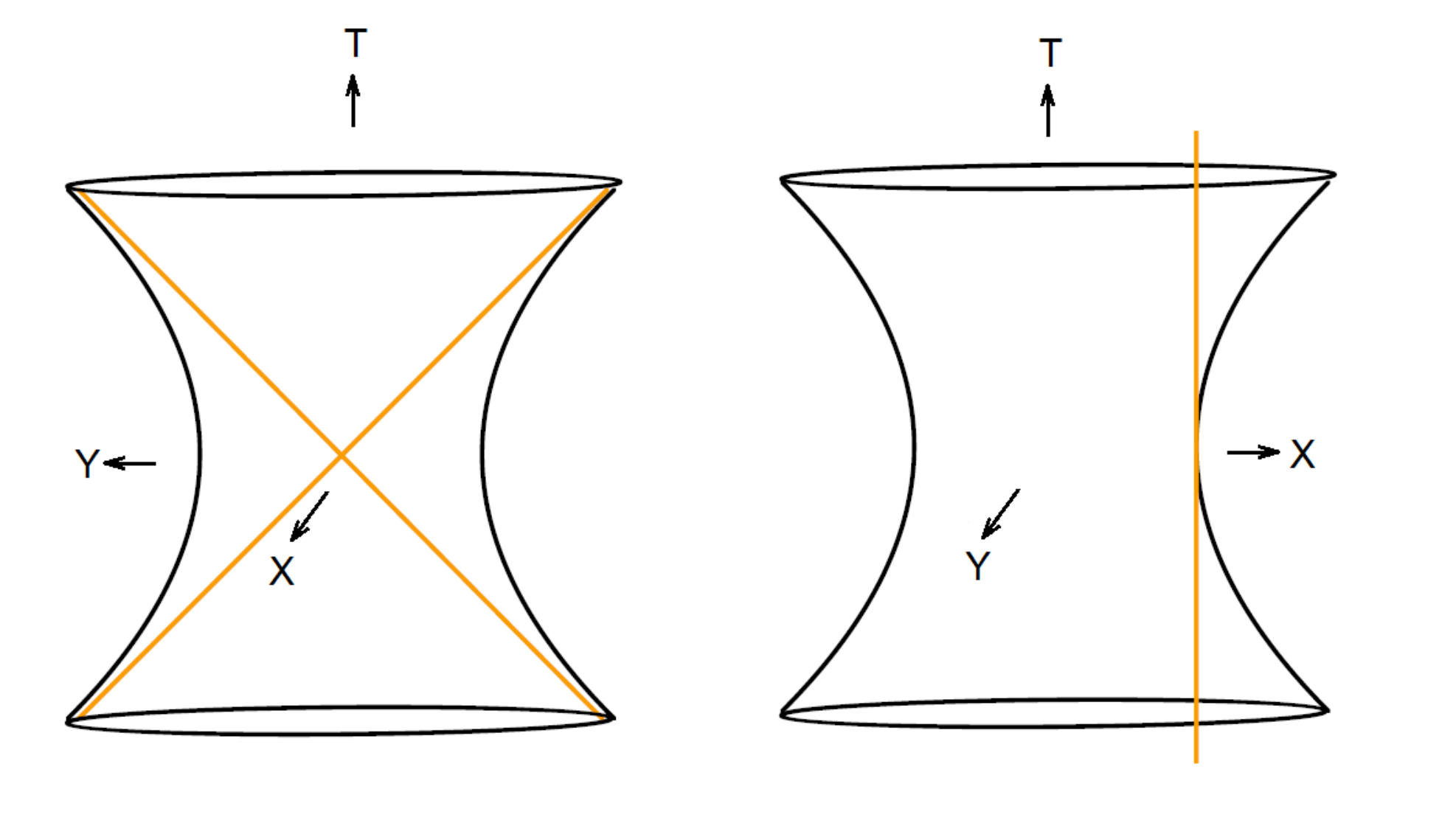}
		\caption{dS$_2$ as a unit hyperboloid in 3D Minkowski space, as seen from two different perspectives. The coordinate $T$ describes the ``height" along the hyperboloid while the coordinates $(X, Y)$ parameterize the circle of radius $\sqrt{1 + T^2}$ at a given $T$. The intersection of the hyperboloid with the plane $X = 1$ (orange) is the horizon of a static patch. Other horizons and static patches are related to this one by actions of the $O(2,1)$ isometry group.}
		\label{dS2}
	\end{center}
\end{figure}

Wick rotating ${T \to -\mathrm{i}Z}$, we see that the Euclidean version of dS$_2$ is simply a 2-sphere, described by the equation 
\begin{equation}
	\vec{X}\cdot\vec{X} = X^2 + Y^2 + Z^2 = \ell^2
	\label{sphere}
\end{equation}
The embedding space for the Euclidean geometry is 3D Euclidean space, on which  
\be 
\vec{X} = (X, Y, Z)
\label{vecX}
\ee
are global Cartesian coordinates.

As we will explain at the end of this subsection, we will ultimately be interested in a solution to \eqref{R=2} whose spacetime is a certain bounded submanifold of dS$_2$. For now, however, in order to keep building necessary tools and intuition, we will continue working with the full dS$_2$ manifold as our (temporary) solution to \eqref{R=2}.

The metric equation of motion, obtained by varying \eqref{action} with respect to $g_{ab}$, sets the dilaton profile to be a solution of
\be 
0 = \inp{\nabla_a\nabla_b-g_{ab}\nabla^2-g_{ab}}\phi
\label{metricEOM}
\ee
where $\nabla_a$ is the covariant derivative operator (Levi-Civita connection) of the 2D metric $g_{ab}$. Maldacena, Turiaci, and Yang \cite{Maldacena:2019cbz} (as well as \cite{Cotler:2019nbi} and subsequent works) considered a solution of \eqref{metricEOM} in which the dilaton is proportional to the timelike embedding space coordinate $T$
\be 
\Phi_{\mathrm{MTY}} = \frac{\Phi_0T}{\ell}
\label{fi=T}
\ee
In this solution, there is no frame in which the dilaton is static. This does not suit our purpose, which is to study static patch holography.

There is another family of solutions in which the dilaton is proportional to a spacelike\footnote{Here by ``spacelike" we mean that the continuation to Lorenzian signature is spacelike.} direction. A particular example is,
\be 
\Phi = \frac{\Phi_0X}{\ell}
\label{fi=X}
\ee
with $\Phi_0 > 0$ some overall positive constant. It is (an appropriate restriction of) this solution that we will consider in what follows, and which we conjecture to be dual to the thermofield double state of $\mathrm{DSSYK}_{\infty}$. Such solutions, and their connection to dimensional reductions of dS$_3$, were previously studied in \cite{Svesko:2022txo}.

\begin{figure}[H]
	\begin{center}
		\includegraphics[scale=.375]{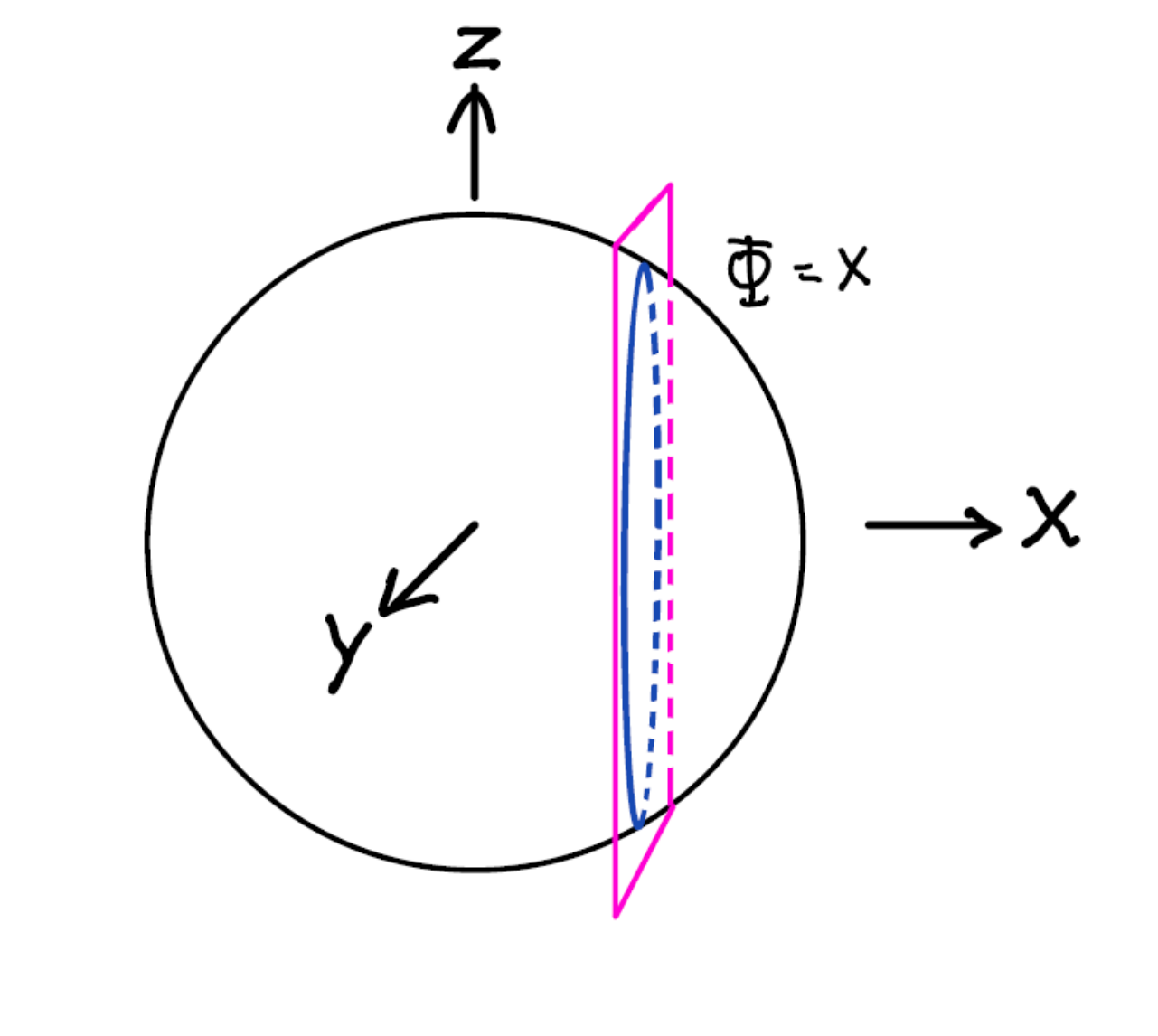}
		\caption{A plane of constant $X$ intersects the sphere on a circle of constant dilaton.}
		\label{2sphere}
	\end{center}
\end{figure}

The various possible solutions of \eqref{metricEOM} can be understood in a unified way by first considering the Euclidean theory, in which the geometry is a 2-sphere \eqref{sphere}. Consider the solution \eqref{fi=X} and any plane $X=a$ (with $|a| \leq \ell$). The dilaton will be constant on the circle where this plane intersects the sphere (see figure \ref{2sphere}). 

We can use the initial solution %\eqref{fi=phi0X} 
\eqref{fi=X} to generate a whole family of solutions by acting with the $O(3)$ isometry group of the sphere (the Euclidean analog of \eqref{dSIso}). This family of solutions can be parameterized by a unit normal vector $\vec{n}$ and the overall normalization constant $\Phi_0$ as
\be
\Phi_{\vec{n}} = \frac{\Phi_0\,\vec{n} \cdot \vec{X}}{\ell}
\label{fi=ndotX}
\ee
The %rotation 
$O(3)$ symmetry of the sphere is spontaneously broken to $U(1)$ by any particular choice of $\vec{n}$, i.e. by any particular choice of dilaton profile from the family of solutions \eqref{fi=ndotX}.

In the solution \eqref{fi=T} of \cite{Maldacena:2019cbz,Cotler:2019nbi}, the unit vector $\vec{n}$ is chosen to lie along the Euclidean time direction $Z$, which is then continued ${Z\to \mathrm{i} T}$ to obtain the corresponding Lorenzian solution.  Both \cite{Maldacena:2019cbz} and \cite{Cotler:2019nbi} take the %constant $\Phi_g$ in \eqref{fi=T} to be real in Lorentz signature, so that the 
dilaton profile \eqref{fi=T} %is 
to be purely real in Lorenzian signature, which forces the the dilaton profile \eqref{fi=T} to be purely imaginary in Euclidean signature \cite{Maldacena:2019cbz}
\be 
\Phi_{\mathrm{MTY}}^{(\mathrm{Euclidean})} = -\mathrm{i}\,\frac{\Phi_0Z}{\ell}
\label{phiMTYEuclidean}
\ee
%It is ultimately is the origin of the imaginary value of the dilaton in that solution. 
This is appropriate for thinking of the dilaton as a clock, but is not appropriate for our purposes since we would like to think of the dilaton as a measure of the degrees of freedom in our system (see subsection \ref{entropy}) or, more specifically, as a spatial compactification scale (see subsection \ref{3D}).

\begin{figure}[H]
	\begin{center}
		\includegraphics[scale=.325]{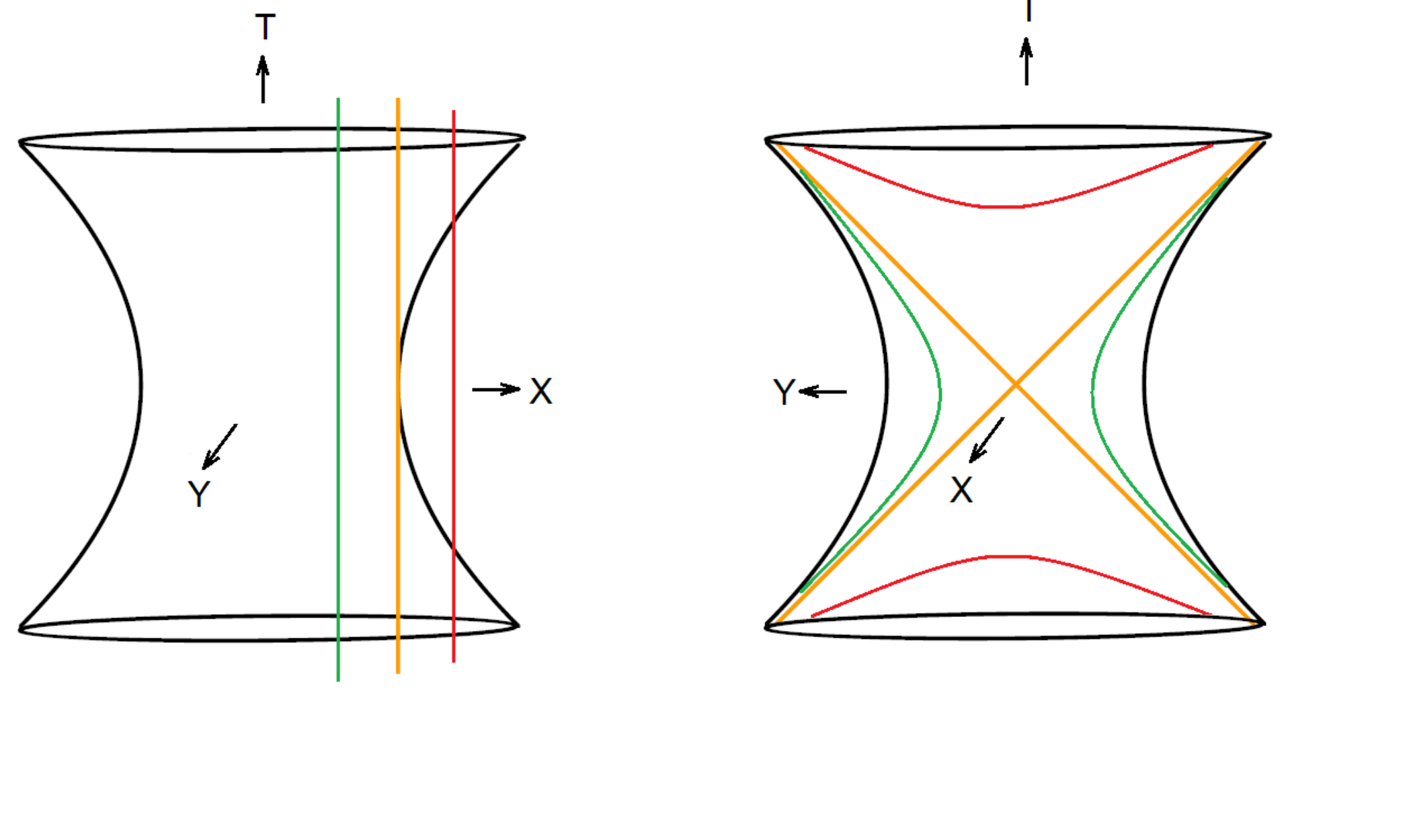}
		\caption{The dS$_2$ hyperboloid and its intersection with various planes of constant $X$ as seen from two perspectives (``from the side" and ``head on"). The orange plane is at $X=1$; its intersection with the hyperboloid forms two null lines (orange) which are the horizons of a static patch. For $|X|<1$ the intersections are timelike and shown in green.  For $|X|>1$ they are spacelike and shown in red.} 
		\label{hype2}
	\end{center}
\end{figure}

In our solution the unit vector $\vec{n} $ is chosen in the spacelike direction $X$. After continuing back to Lorentz signature, the profile of the dilaton can be visualized by considering the intersection of the dS$_2$ hyperboloid with various planes of constant $X$, see figure \ref{hype2}.

%
%\begin{figure}[H]
%\begin{center}
%\includegraphics[scale=.3]{hyperboloid}
%\caption{Unit hyperboloid and intersecting plane seen from two angles. The %plane is at $X=1.$ The intersection with the hyperboloid is two light-like lines %which are the horizons of a static patch. For $X<1$ the intersections are time-%like and for $x>1$ they are space-like.} 
%\label{hyperboloid}
%\end{center}
%\end{figure}
%

%
%

%Next let us consider a static patch of dS$_2$. In static patch coordinates $(t,r)$, the metric takes the standard form,
The two regions of the dS$_2$ hyperboloid in which the lines of constant  $\Phi = \Phi_0X/\ell$ are everywhere timelike constitute a pair of static patches (i.e. a pair of non-intersecting maximal size causal patches in dS). Following \cite{Susskind:2021omt}, we will call one of them the ``pode" static patch and the other the ``antipode" static patch, and we will call their respective null boundaries the ``pode" and ``antipode" horizons.

Consider the e.g. pode static patch. In this region, we can erect (pode) static patch coordinates $(t, r)$ in which the dS$_2$ metric takes the familiar form
\be 
\mathrm{d}s^2 = -\inp{1-\frac{r^2}{\ell^2}}\mathrm{d}t^2 + \inp{1-\frac{r^2}{\ell^2}}^{-1}\mathrm{d}r^2
\label{metric}
\ee
We take $t$, which runs from $-\infty$ to $+\infty$, to increase in the direction of increasing $T$. As its name suggests, the static patch has a timelike ``boost" symmetry generated by the Killing vector $\partial/\partial t$. The continuation of this Killing vector to the antipode static patch then points in the direction of decreasing $T$. In order to clean up various formulae, it is helpful to introduce the shorthand 
\be 
f(r) \equiv 1-\frac{r^2}{\ell^2}
\label{f(r)}
\ee

\begin{figure}[H]
	\begin{center}
		\includegraphics[scale=1.1]{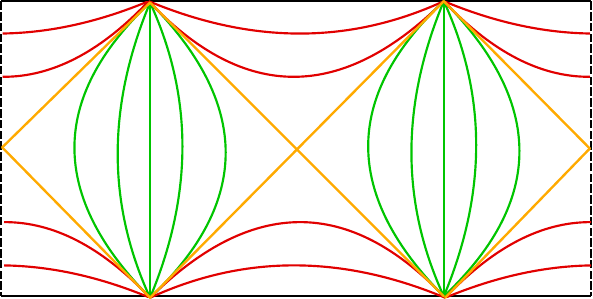}
		\caption{Conformal diagram of dS$_2$ showing the lines of constant dilaton and the two antipodal static patches (regions where the surfaces of constant dilaton are timelike/green). The diagram should be understood to be periodic in the horizontal direction.}
		\label{conf}
	\end{center}
\end{figure}

Figure \ref{conf} shows the full conformal diagram of full dS$_2$ with the two antipodal static patches and the lines of constant dilaton indicated with the same color coding as in figure \ref{hype2}. Note that we must identify the left and right edges of the conformal diagram, which should be understood to be periodic in the horizontal direction. To be concrete, we will regard the right static patch in figure \ref{conf} to be the ``pode" static patch, and the left static patch in figure \ref{conf} to be the ``antipode" static patch.

In full dS$_2$---unlike for dS$_{D\geq 3}$---the coordinate $r$ in \eqref{metric} runs from $-\ell$ to $+\ell$ from right to left within the pode static patch (and the opposite way within the antipode static patch). In dS$_{D\geq 3}$, by comparison, the coordinate $r$ only runs from $0$ to $\ell$. This is why the conformal diagram of full dS$_2$ (figure \ref{conf}) is twice as wide as the Penrose diagram (which coincides with the conformal diagram for dS$_{D\geq 3}$). We will return to this point when we discuss our solution of interest \eqref{dS2'} at the end of this subsection.

The cosmological horizon of the static patch is at $|r| = \ell$. In what follows, we will consider a timelike ``stretched" horizon at $|r| = r_{\mathrm{sh}}$ with $1 - \frac{r_{\mathrm{sh}}}{\ell} \ll 1$. More specifically, we will take $r_{\mathrm{sh}}$ to lie one Planck length\footnote{As mentioned in a previous footnote, the only place in this paper where the precise location of the stretched horizon plays a role is in the derivation of the timescale \eqref{tscr} controlling bulk emergence from the one-sided retarded two-point function. This timescale is preserved as long as the stretched horizon is placed a distance from the mathematical horizon which is polynomial in the Planck length, $\delta s/\ell \ \sim \ \inp{\ell_{\mathrm{P}}/\ell}^n$, $n \geq 1$.} in proper distance from the horizon, so that
\be
1-\frac{r_{\mathrm{sh}}}{\ell} = \frac{1}{2}\inp{\frac{\ell_{\mathrm{P}}}{\ell}}^2\times\Big(1 + O\big(\ell_{\mathrm{P}}^2/\ell^2\big)\Big)
\label{rsh}
\ee

There are of course many pairs of static patches in dS$_2$, related to one another by actions of the $O(2,1)$ symmetry group, but---apart from a trivial ambiguity\footnote{Namely, up to an action of the $\partial/\partial t$ Killing symmetry and/or a $\mathbb{Z}_2$ exchange of the pode and antipode (including a flip in the time orientation). These are the 2D versions of the ``easy" symmetries of \cite{Susskind:2021omt} and of $\mathsf{CPT}$, respectively.}---only one of them has the property that, in it, the dilaton \eqref{fi=X} is static. In that frame $X = r$ and so the dilaton is given by
\be 
\Phi = \frac{\Phi_0r}{\ell}
\label{fi=r}
\ee
In this way a choice of spacelike unit vector $\vec{n}$ in \eqref{fi=ndotX} is equivalent to a choice of a pair of pode and antipode static patches in dS$_2$.

We now come to our desired solution---which we will refer to as dS$_2{}'$---that was teased earlier. The origin of this solution as a dimensional reduction of dS$_3$ will be explained in section \ref{3D} below. For now we introduce it as an interesting solution in its own right. The construction is as follows: Once we have chosen a pair of static patches (i.e. once we have made a choice of the spacelike unit vector $\vec{n}$ in \eqref{fi=ndotX}) and erected pode static patch coordinates $(t,r)$, we can define a corresponding choice of global coordinates $(\tau, x)$ which extend to the whole dS$_2$ hyperboloid via 
\begin{align}
	\sinh(\tau/\ell) &= \inp{1-\frac{r^2}{\ell^2}}^{1/2}\sinh(t/\ell)
	%\sinh(\tau/\ell) &= \sqrt{f(r)}\,\sinh(t/\ell)
	\label{tglobal} \\
	\bn \nonumber\\
	\ell\cos(x) &= \frac{r}{\sqrt{1 + \inp{1-\frac{r^2}{\ell^2}}\sinh^2(t/\ell)}}
	%\ell\cos(x) &= \frac{r}{\sqrt{1 + f(r)\sinh^2(t/\ell)}}
	\label{rglobal}
\end{align}
The global time coordinate $\tau$ runs from $-\infty$ to $+\infty$ from the bottom to the top of the conformal diagram while $x$---which parameterizes the spatial circle---runs from $-\pi$ to $+\pi$ from right to left on the conformal diagram (with of course $x \sim x + 2\pi$). In terms of these coordinates, the metric \eqref{metric} and dilaton \eqref{fi=r} can be extended to the whole conformal diagram/dS$_2$ hyperboloid via 
\begin{align}
	\mathrm{d}s^2 &= -\mathrm{d}\tau^2 + \ell^2\cosh^2(\tau/\ell)\,\mathrm{d}x^2
	\label{metricglobal}\\
	\Phi &= \cosh(\tau/\ell)\cos(x)
	\label{figlobal}
\end{align} 

\begin{figure}[H]
	\begin{center}
		\includegraphics[scale=1.1]{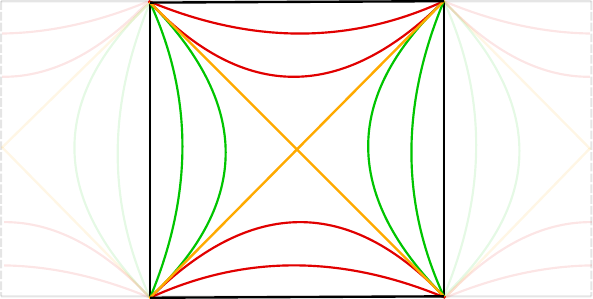}
		\caption{Conformal diagram of dS$_2{}'$, viewed as a bounded submanifold of dS$_2$.}
		\label{restricted}
	\end{center}
\end{figure}

\begin{figure}[H]
	\begin{center}
		\scalebox{0.75}{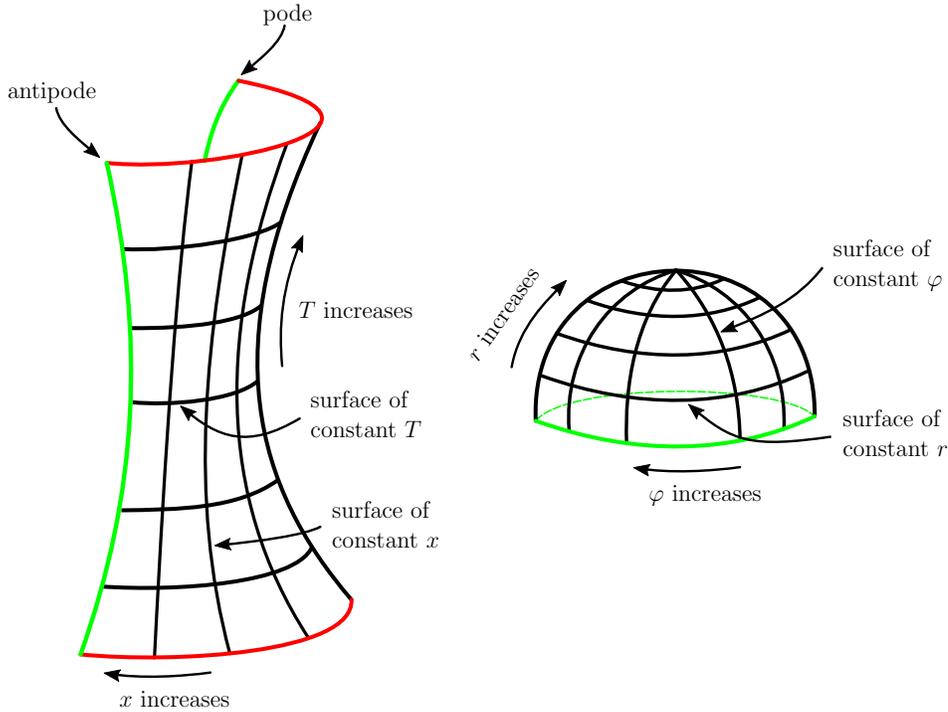}
		\caption{Lorentzian dS$_2{}'$ (left) embedded in (2+1)-dimensional Minkowski space and Euclidean dS$_2{}'$ (right) embedded in 3D Euclidean space. In both figures, the spatial boundary is marked in green.}
		\label{dS2prime}
	\end{center}
\end{figure}

Our desired spacetime, dS$_2{}'$, is the submanifold 
\be 
x \in \insb{-\frac{\pi}{2},+\frac{\pi}{2}}
\label{restriction}
\ee 
of dS$_2$, which leads to the condition
\be 
r \geq 0
\label{restrictionr}
\ee
The conformal diagram of dS$_2{}'$, viewed as a bounded submanifold of full dS$_2$, is shown in figure \ref{restricted}. The embedding of dS$_2{}'$ into $(2+1)$-dimensional Minkowski space is shown in figure \ref{dS2prime}. 

The metric and dilaton of dS$_2{}'$ are the restrictions of \eqref{metric}/\eqref{metricglobal} and \eqref{fi=r}/\eqref{figlobal} to this submanifold respectively. The conformal diagram (figure \ref{restricted}) closely resembles the Penrose diagram of dS$_{D}$ with the key difference being that, in figure \ref{restricted}, slices of constant $T$ are intervals \eqref{restriction} rather than $(D-1)$-spheres. There are two spatial boundaries---one at the pode and another at the antipode---on which we impose Dirichlet boundary conditions for the dilaton.

To summarize, the solution to \eqref{R=2} that we will consider in this paper is the dS$_2$ metric restricted to the interval \eqref{restricted}
\be
-\frac{\pi}{2} \leq x \leq + \frac{\pi}{2} \quad \implies \quad 0 \leq r < \infty
\label{dS2'}
\ee 
and we will call this solution dS$_2{}'$.

The Euclidean version of dS$_2{}'$ is obtained by continuing
\be 
t \to -\mathrm{i}\ell\varphi
\ee 
in \eqref{metric} to obtain 
\be 
\frac{\mathrm{d}s^2_{\mathrm{E}}}{\ell^2} = \inp{1-\frac{r^2}{\ell^2}}\mathrm{d}\varphi^2 +\inp{1-\frac{r^2}{\ell^2}}^{-1}\frac{\mathrm{d}r^2}{\ell^2}
\label{Euclideanmetric}
\ee
Defining $\theta$ via
\be 
\cos(\theta) = \frac{r}{\ell}
\label{r=cost}
\ee
\eqref{Euclideanmetric} becomes 
\be 
\frac{\mathrm{d}s^2_{\mathrm{E}}}{\ell^2} = \sin^2(\theta)\,\mathrm{d}\varphi^2 + \mathrm{d}\theta^2
\label{Emetric}
\ee
This is the metric of a round two-sphere of radius $\ell$, as expected. The restriction $r \geq 0$ amounts to the restriction
\be 
r \geq 0 \ \implies \ \theta \leq \frac{\pi}{2}
\label{S2'}
\ee 
We see that the Euclidean version of the spacetime \eqref{metric}, \eqref{dS2'} is a hemisphere (see figure \ref{dS2prime}). We impose Dirichlet boundary conditions for the dilaton on the bounding circle. 

\subsection{Temperature and Tomperature}
\quad \ dS$_2{}'$ still has the timelike ``boost" Killing symmetry generated by $\partial/\partial t$ (since points on dS$_2{}'$ remain on dS$_2{}'$ under actions of the isometry) for which the cosmic horizon is a bifurcate Killing horizon. This continues to an azimuthal ``rotation" symmetry of Euclidean dS$_2{}'$ generated by $\partial/\partial \varphi$, for which the Euclidean cosmic horizon (which is simply the point $\theta = 0$) is a fixed point. The periodicity 
\be 
\varphi \sim \varphi + 2\pi
\label{period}
\ee 
implies that, semiclassically, the cosmic horizon of dS$_2{}'$ appears to radiate at a temperature given by the Gibbons-Hawking temperature \cite{Gibbons:1977mu}
\be 
T_{\mathrm{GH}} \equiv \frac{1}{\beta_{\mathrm{GH}}} = \frac{1}{2\pi \ell}
\label{TdS}
\ee

The bulk tomperature $\mathcal{T}$ is defined, in the context of static patch horizon holography, as the change in the energy when one qubit of information is emitted from the horizon \cite{Lin:2022nss}. The analogy with the definition given for $\mathrm{DSSYK}_{\infty}$ in section \ref{hightemp} should be clear. As explained in \cite{Lin:2022nss}, this is the same as the Gibbons-Hawking temperature: $\mathcal{T} = T_{\mathrm{GH}}$. We see that in the dS$_2{}'$ \textit{bulk}, the tomperature and semiclassical (Gibbons-Hawking) temperature coincide, and in what follows we will use $\mathcal{T}$ and $T_{\mathrm{GH}}$ (and hence $\mathcal{T}^{-1}$ and $\beta_{\mathrm{GH}}$) interchangeably. Note that this Gibbons-Hawking temperature is \textit{not} the same as the fundamental temperature $T = \beta^{-1}$ appearing in the Boltzmann distribution/thermofield double state \eqref{rhobeta}/\eqref{TFD}, which we have taken to infinity.

\subsection{Entropy}
\label{entropy}
\quad \ We would like to justify our interpretation of the dilaton as a measure of the degrees of freedom in our system. In diffeomorphism invariant theories such as \eqref{action}, there is a notion of entropy that can be ascribed to stationary spacetimes with bifurcate Killing horizons known as the Wald entropy \cite{Wald:1993nt}, which can be thought of as generalizing the notions of the Bekenstein-Hawking entropy \cite{Bekenstein:1973ur,Hawking:1975vcx} for black holes and the Gibbons-Hawking entropy \cite{Gibbons:1977mu} for de Sitter space to theories beyond vacuum Einstein gravity.

It was shown in \cite{Svesko:2022txo} that the Wald entropy of \eqref{action} on the solution \eqref{dS2'} is given by 
\be 
S = \frac{\Phi_0}{4G_N^{(2)}}
\label{SWald}
\ee
In the context of static patch horizon holography we will want to think of this horizon entropy as reflecting the entropy of the holographically dual finite quantum-mechanical system (in our case $\mathrm{DSSYK}_{\infty}$ with $S = \lfloor N \rfloor /2$). It is in this sense that we consider the dilaton \eqref{fi=X} to be a measure of the degrees of freedom in our system. As we will explain in section \ref{3D} below, we will be motivated to make the choice $\Phi_0 = 1$, so that 
\be 
S = \frac{2\pi\ell}{4\ell_{\mathrm{P}}}
\label{S}
\ee
where $\ell_{\mathrm{P}}$ was defined in \eqref{ellP} above. In terms of the entropy $S$, we have
\be 
1-\frac{r_{\mathrm{sh}}}{\ell} = \frac{\pi^2}{8S^2}\,\times\inp{1 + O(S^{-2})}
\label{rshvsS}
\ee 
where $r_{\mathrm{sh}}$ was defined in \eqref{rsh} above.

\subsection{Relation to Three Dimensional de Sitter }
\label{3D}
\quad \ We would now like to motivate the restriction \eqref{restriction} that gave rise to our solution \eqref{metric},\eqref{fi=r},\eqref{dS2'} of \eqref{action}.

The solution \eqref{metric},\eqref{fi=r},\eqref{dS2'} can be understood as a dimensional reduction of three-dimensional de Sitter space dS$_3$. We can see this as follows: Begin by picking a static patch in dS$_3$. In this patch we can erect 3D static patch coordinates $(t, r, \alpha)$ in which the metric takes the form
\be
%\mathrm{d}s^2_{\mathrm{dS}_3}=-\inp{1-\frac{r^2}{\ell^2}}\mathrm{d}t^2 + \inp{1-\frac{r^2}{\ell^2}}^{-1}\mathrm{d}r^2 +r^2 \mathrm{d}\alpha^2
\mathrm{d}s^2_{\mathrm{dS}_3}=-f(r)\,\mathrm{d}t^2 + \frac{\mathrm{d}r^2}{f(r)} +r^2 \mathrm{d}\alpha^2
\label{dS3}
\ee
with $f(r)$ defined as in \eqref{f(r)} above. Here $\alpha$ is an angle running from $0$ to $2\pi$ which coordinatizes the circles of constant $(t, r)$. The geometry is circularly-symmetric since the metric is independent of $\alpha$; in other words, there is a $U(1)$ symmetry in the angular direction along each circle of fixed $(t,r)$. The coordinate $r$ now runs from $0$ to $\ell$ from right to left within the pode static patch (and the opposite way within the antipode static patch), as is usually the case in dS$_{D\geq 3}$.

As above, we can define a corresponding set of global coordinates $(\tau, x, \alpha)$---which extend to the entire dS$_3$ manifold---via \eqref{tglobal}, \eqref{rglobal} and $\alpha = \alpha$. The coordinate $x$ now only runs from $-\pi/2$ to $+\pi/2$. The reason for the differing range of $x$ can be understood as follows: The metric of dS$_D$ can be written globally---in any spacetime dimension---as
\be 
\mathrm{d}s^2 = -\mathrm{d}\tau^2 + \ell^2\cosh^2(\tau/\ell)\,\mathrm{d}\Omega_{(D-1)}^2
\label{dsdOmega}
\ee 
with $\tau$ as in \eqref{tglobal}. The differing ranges of $x$ boil down to the simple fact that 
\be 
\mathrm{d}\Omega_1^2 = \mathrm{d}x^2 
\label{dOmega2}
\ee 
with $x \in [-\pi,\pi]$ parameterizing the whole spatial $\mathbb{S}^1$, while 
\be 
\mathrm{d}\Omega_{D\geq 2} = \mathrm{d}x^2 + \cos^2(x)\,\mathrm{d}\Omega_{(D-1)}^2
\label{dOmega}
\ee 
with $x \in [-\frac{\pi}{2},+\frac{\pi}{2}\,]$ parameterizing a semicircle running between the poles of the spatial $\mathbb{S}^{D\geq2}$.

We now recall that the 2D JT gravity action \eqref{action} can be obtained by dimensional reduction of the 3D Einstein-Hilbert action\footnote{Here $\mathsf{x}$ is a 3D coordinate on the 3D spacetime $\mathcal{M}^{(3)}$ with 3D metric $\mathsf{g}_{\mu\nu}$.} with positive cosmological constant $\Lambda_{(\mathrm{3D})} = 1/\ell^2$ \cite{Svesko:2022txo}:
\be
I_{\mathrm{EH}}[\mathsf{g}] = \frac{1}{16\pi G_N^{(3)}}\int_{\mathcal{M}^{(3)}}\mathrm{d}^3\mathsf{x}\sqrt{|\mathsf{g}|}\inp{\mathsf{R}-2\Lambda_{(\mathrm{3D})}}
\label{3DCC}
\ee
The reduction proceeds by restricting to spacetime manifolds of the form $\mathcal{M}^{(3)} = \mathcal{M}^{(2)}\times \mathbb{S}^1$ and parameterizing the 3D line element in terms of a 2D metric $g_{ab}$ and a dilaton $\Phi$---both independent of the angular direction along the $\mathbb{S}^1$---via
\be 
\mathrm{d}s^2_{(\mathrm{3D})} = g_{ab}(x)\,\mathrm{d}x^a\,\mathrm{d}x^b + \ell^2\,\Phi^2(x)\,\mathrm{d}\alpha^2
\label{dimredansatz}
\ee
Here $x$ is a 2D coordinate on $\mathcal{M}^{(2)}$ and $\alpha$ is, as above, an angular coordinate on the $\mathbb{S}^1$. One can check that the action for the dimensionally reduced theory for $g_{ab}$ and $\Phi$ is given by \eqref{action} (see e.g. \cite{Svesko:2022txo} for details). The 2D and 3D de Sitter radii $\ell$ coincide, while the 2D and 3D Newton constants are related by 
\be 
\frac{1}{G_N^{(2)}} = \frac{2\pi\ell}{G_N^{(3)}}
\label{GN}
\ee 
We see that the definition \eqref{ellP} amounts to demanding that the 2D and 3D Planck lengths also coincide. 

We now rewrite \eqref{dS3} in the form
\bea
\mathrm{d}s^2 %\eq -\inp{1-\frac{r^2}{\ell^2}}\mathrm{d}t^2 +\inp{1-\frac{r^2}{\ell^2}}^{-1}\mathrm{d}r^2 +\ell^2\,\Phi^2 \mathrm{d}\alpha^2 \cr
\eq -f(r)\,\mathrm{d}t^2 +\frac{\mathrm{d}r^2}{f(r)} +\ell^2\Phi^2 \mathrm{d}\alpha^2 \cr
\Phi \eq \frac{r}{\ell}
\label{alpha}
\eea
with, again, $0 \leq r < \ell$. This corresponds to the solution \eqref{metric},\eqref{fi=r} but with $x$ restricted to the range \eqref{restriction}. In other words, this corresponds to the solution dS$_{2}'$ \eqref{metric},\eqref{fi=r},\eqref{dS2'} from the previous subsection, which we now see is simply a dimensional reduction of dS$_3$ (\footnote{From this perspective, another way to view the origin of the restriction \eqref{dS2'} comes from the fact that the locus $r = 0$ in dS$_2{}'$ descends from a polar coordinate singularity of dS$_3$.}); the dilaton is simply the compactification scale of the $\mathbb{S}^1$ in de Sitter units. 

From the perspective of the dimensional reduction, the restriction \eqref{dS2'} is fundamental: we cannot extend the dimensional reduction of the solution \eqref{alpha} to the full dS$_2$ manifold without either letting the dilaton become negative---and therefore losing its interpretation as a compactification scale---or extending $\Phi$ as $\Phi = \ell^{-1}|r|$ and violating the equation of motion \eqref{metricEOM} at $r = 0$, where the proposed dilaton profile $\Phi = \ell^{-1}|r|$ is nondifferentiable.

Semiclassically, we can associate a Gibbons-Hawking entropy $S_{\mathrm{GH}}$ to the cosmological horizon of \eqref{dS3} (viewed as a solution of \eqref{3DCC}) in the usual way via \cite{Gibbons:1977mu}
\be
S_{\mathrm{GH}} = \frac{A^{(3)}}{4G_N^{(3)}} = \frac{2\pi \ell}{4\ell_{\mathrm{P}}}
\label{S3D}
\ee
\eqref{S3D} coincides with the Wald entropy \eqref{S} provided we take $\Phi_0 = 1$.

\subsection{Remark on Symmetry Breaking}

\quad \ A key question \cite{Goheer:2002vf,Susskind:2021omt,Chandrasekaran:2022cip} in static patch horizon holography regards the implementation of the de Sitter symmetries \eqref{dSIso}. It is commonly thought that these symmetries should be regarded---at least in the context of pure de Sitter space and fluctuations thereof in Einstein gravity---as approximate symmetries emerging at large $S$ which are broken by exponentially small \cite{Susskind:2021omt} finite $S$ effects \cite{Goheer:2002vf}. However in the context of our model \eqref{action},\eqref{metricglobal},\eqref{figlobal},\eqref{dS2'} the situation is even more severe: the symmetry operations do not exist even at the classical level.

Indeed the 2D parent solution \eqref{metric}, \eqref{fi=r} is already not invariant under the full de Sitter isometry group \eqref{dSIso}, which is $O(2,1)$ in two spacetime dimensions. The metric itself is exactly globally dS$_2$, but the choice of vector $\vec{n}$ in \eqref{fi=ndotX} spontaneously breaks the symmetry\footnote{Here $(\mathrm{discrete})$ includes the discrete symmetries of parity $\mathsf{P}$, time reversal $\mathsf{T}$, and their product $\mathsf{PT}$. It is possible that there may be further anomalies breaking these symmetries due to e.g. bulk matter content, additional topological terms in the bulk action, and/or configuration space restrictions in the gravitational path integral, as can happen in AdS JT gravity \cite{Stanford:2019vob}.} $O(2,1) \to \mathbb{R}\times(\mathrm{discrete})$. The situation with our dS$_2{}'$ solution \eqref{dS2'} is even more extreme, since the spacetime itself also breaks the symmetry (in fact the spacetime itself only carries a representation of $\mathbb{R} \times(\mathrm{discrete}) \subset O(2,1)$).

The situation is a bit clearer from the perspective of the 3D parent solution \eqref{dS3}. Classically (i.e. on the solution \eqref{dS3}), the $(2+1)$-dimensional parent theory \eqref{3DCC} has an exact $O(3,1)$ symmetry. Semiclassically, the $O(3,1)$ symmetry should be regarded as a gauge symmetry (see \cite{Susskind:2021omt}) which transforms one static patch into another; since it is a gauge symmetry, we are free to fix the gauge\footnote{More specifically, it is the ``hard" symmetries of \cite{Susskind:2021omt} which exchange different static patches. Once we have fixed these symmetries by specifying a pair of static patches, there is still a subgroup of $SO(D,1)$---which \cite{Susskind:2021omt} calls the ``easy" symmetries---which act within a given static patch by boosts generated by $\partial/\partial t$ and rotations in the angular direction(s).} by making a particular choice of static patch.

When viewed through the lens of dimensional reduction, the lack of symmetry in the dS$_2{}'$ solution \eqref{dS2'} originates from the fact that it is obtained from the \textit{gauge-fixed} version of the 3D parent solution. In particular, when performing the dimensional reduction \eqref{alpha} we must first choose a particular 3D static patch. This brings to light a possible complication: The $O(3,1)$  symmetry of dS$_3$ will be anomalous if it is to be realized by a system of finite entropy \cite{Goheer:2002vf}---it must be explicitly broken by exponentially small effects \cite{Susskind:2021omt}. What this implies for the dimensional reduction \eqref{alpha} when these exponentially small effects become relevant is unclear, but the issue is largely irrelevant in the semiclassical limit $S \to \infty$.

\section{Correlation Functions and Bulk Emergence}
\label{2ptonesided}
\quad \ We begin our study of the dS-JT/$\mathrm{DSSYK}_{\infty}$ dictionary by showing that we can reproduce the behavior \eqref{G0(t)} using quasinormal modes in the proposed bulk dual \eqref{metric},\eqref{fi=r},\eqref{dS2'}. We then go on to describe some early-time features of the bulk one-sided retarded two-point function \eqref{t0} for propagating bulk matter fields. This correlation function provides an interesting example of bulk emergence which is governed by the timescale $t_* \sim \beta_{\mathrm{GH}}\log(S)$ which would have been the scrambling time in an ordinary fast-scrambling theory. This correlation function also places some constraints on the boundary-to-bulk operator mapping, which we will discuss in \eqref{bb} below.

\subsection{Bulk Correlators and Quasinormal Modes}
\quad \ We might expect $\mathrm{DSSYK}_{\infty}$ correlation functions like \eqref{G(t)beta} to encode the boundary (i.e. horizon) extrapolates of correlation functions of bulk fields. In higher dimensions, this correspondence might be complicated by the fact that the boundary theory is likely not itself a local quantum field theory, but in two dimensions, where the horizons are (at any given horizon time) just points with no additional stucture, one may hope that correlators of simple bulk operators restricted to the horizon may be related in a simple way to correlators of simple operators in the holographically dual boundary theory, namely $\mathrm{DSSYK}_{\infty}$. In this section we will test that expectation and see what we find.

While we would certainly expect that---should they exist---the bulk fields which would reproduce \eqref{G0(t)} would be fermionic, we will consider here for simplicity a real scalar field\footnote{Another reason to consider a real scalar field is that, according to \cite{Du:2004jt}, there are no fermionic quasinormal modes in the $s$-wave sector in dS$_3$. We will explain how this may actually be a feature---rather than a bug---of the conjectured dS-JT/$\mathrm{DSSYK}_{\infty}$ duality in section \ref{bb} below.}. The leading decaying quasinormal modes of different bulk fields may differ by an $O(\ell^0)$ numerical factor, so we will only seek to reproduce \eqref{G0(t)} up to an overall $O(\mathcal{T}^0)$ numerical factor in the exponent. 

One might expect that the simplest and most natural object to study would be a light scalar field minimally coupled to the metric but not to the dilaton of \eqref{action}. This would be a perfectly natural object to study from an intrinsically two-dimensional perspective, but is a rather unnatural object to study from the dimensionally reduced perspective of section \ref{3D}, since such a field would \textit{not} arise from the dimensional reduction of a minimally-coupled scalar field $\upphi$ in three dimensions. The latter case would instead give rise to a 2D field $\phi$ which is coupled to the dilaton $\Phi$ via\footnote{In the more familiar JT gravity with negative cosmological constant obtained from the dimensional reduction of the near-horizon limit of a near-extremal black hole (as well as in the case considered by \cite{Maldacena:2019cbz,Cotler:2019nbi}), the dimensional reduction of a higher dimensional minimally coupled scalar again gives rise---at leading order in the semiclassical expansion---to a 2D minimally coupled scalar. This is because, in that context, one expands the dilaton about a constant value $\Phi(x) = \Phi_0 + S^{-\frac{1}{2}}\,\phi(x) + \dots$. We perform no such expansion here.} 
\begin{align}
	-I_{(\mathrm{3D})}[\upphi] 
	= \ &\frac{1}{2}\int\mathrm{d}^3\mathsf{x}\sqrt{|\mathsf{g}|}\inp{\mathsf{g}^{\mu\nu}\,\nabla_{\mu}\upphi\nabla_{\nu}\upphi + m^2\upphi^2}\label{3Dscalar}\\
	\to \ &\frac{1}{2}\int\mathrm{d}^2x\sqrt{|g|}\,\Phi\inp{g^{ab}\,\nabla_a\phi\nabla_b\phi + m^2\phi^2}
	\label{DCscalar}
\end{align}
We will therefore study a 2D scalar field described by the action \eqref{DCscalar}.

For our purposes, it is enough to work at leading order in the semiclassical expansion, i.e. to put the metric and dilaton on-shell as in \eqref{metric},\eqref{fi=r},\eqref{dS2'} and treat the field $\phi$ as living on a fixed background. The boundary condition for $\phi$ inherited from the 3D parent spacetime \eqref{dS3} is Neumann reflecting, $\partial_r\phi\big|_{r = 0} = 0$. 

We want to calculate the one-sided retarded two-point function $G^{(\mathrm{R})}(t)$ of the field $\phi$, defined by
\be
G^{(\mathrm{R})}(t) 
\equiv \mathrm{i}\,\Theta(t)\,\big\la\,\Omega'\,\big|\,[\phi(t,r_{\mathrm{sh}}),\phi(0,r_{\mathrm{sh}})]\,\big|\,\Omega'\,\big\ra
\label{t0}
\ee
or, equivalently but somewhat more usefully, 
\be 
G^{(\mathrm{R})}(t) 
= \mathrm{i}\,\Theta(t)\,\big\la\,\Omega'\,\big|\,[\phi(+t/2,r_{\mathrm{sh}}),\phi(-t/2,r_{\mathrm{sh}})]\,\big|\,\Omega'\,\big\ra
\label{tt}
\ee
Here $(t,r)$ are static patch coordinates as in \eqref{metric}, $r_{\mathrm{sh}}$ is the radius of the stretched horizon as in \eqref{rsh}, $[\,\cdot\,,\,\cdot\,]$ denotes the commutator, and $|\,\Omega'\,\ra$ denotes the analog\footnote{i.e. it is given by the path integral on the half-hemisphere over all regular (nonsingular) field configurations which obey the Neumann reflecting boundary condition at the ``hemisphere" part of the boundary. It can also be understood as the vacuum state picked out by the boost Killing vector $\partial/\partial t$ of \eqref{metric},\eqref{dS2'} and hence as the dual of the ground state of the boost Hamiltonian $H = H_R-H_L$ which acts on the Hilbert space of the (infinite $N$) thermofield double.} of the Euclidean vacuum/Bunch-Davies-Hartle-Hawking state \cite{Chernikov:1968zm,Bunch:1978yq,Hartle:1983ai} for the spacetime \eqref{metric},\eqref{dS2'} and action \eqref{DCscalar}. The equivalence of \eqref{t0} and \eqref{tt} follows from the boost invariance of the state $|\,\Omega'\,\ra$. Of course, \eqref{t0},\eqref{tt} can be related to the more conventional Wightman function
\be 
G(t) \equiv \la\,\Omega'\,|\,\phi(+t/2,r_{\mathrm{sh}})\,\phi(-t/2,r_{\mathrm{sh}})\,|\,\Omega'\,\ra
\label{wight}
\ee
via\footnote{This simply follows from the fact that
	\be 
	\la\,\phi(x)\,\phi(x')\,\ra^* = \la\,\phi(x')\,\phi(x)\,\ra
	\ee} 
\be
G^{(\mathrm{R})}(t) = -2\,\Theta(t)\,\mathrm{Im}\,G(t)
\label{RvsW}
\ee

The two-point function \eqref{tt} of the field $\phi$ with action \eqref{DCscalar} is difficult to calculate directly, but it should agree with the %one-sided 
retarded two-point function 
\be 
\mathsf{G}^{(\mathrm{R})}_{(\text{$s$-wave})}(t) \equiv \mathrm{i}\,\Theta(t)\,\big\la\,\Omega\,\big|\,[\upphi_0(+t/2,r_{\mathrm{sh}}),\upphi_0(-t/2,r_{\mathrm{sh}})]\,\big|\,\Omega\,\big\ra
\ee 
of the $s$-wave mode $\upphi_0$ of the 3D minimally-coupled scalar field $\upphi$ in the dS$_3$ Euclidean vacuum $|\,\Omega\,\ra$. In other words, we expect that
\begin{align}
	G^{(\mathrm{R})}(t) %\mathsf{G}^{(\mathrm{R})}_{(\text{$s$-wave})}(t)
	&%\equiv 
	=
	\mathrm{i}\,\Theta(t)\,\big\la\,\Omega\,\big|\,[\upphi_0(+t/2,r_{\mathrm{sh}}),\upphi_0(-t/2,r_{\mathrm{sh}})]\,\big|\,\Omega\,\big\ra\\
	&= \mathrm{i}\,\Theta(t)\times\frac{1}{4\pi^2}\intf{0}{2\pi}\mathrm{d}\alpha\,\mathrm{d}\alpha'\,\big\la\,\Omega\,\big|\,[\upphi(+t/2,r_{\mathrm{sh}},\alpha),\upphi(-t/2,r_{\mathrm{sh}},\alpha')]\,\big|\,\Omega\,\big\ra\\
	%&= \frac{\Theta(t)}{4\pi^2}\intf{0}{2\pi}\mathrm{d}\alpha\,\mathrm{d}\alpha'\,\big\la\,\Omega\,\big|\,[\upphi(+t/2,r_{\mathrm{sh}},\alpha-\alpha'),\upphi(-t/2,r_{\mathrm{sh}},0)]\,\big|\,\Omega\,\big\ra\\
	&= \frac{1}{4\pi^2}\intf{0}{2\pi}\mathrm{d}\alpha\,\mathrm{d}\alpha'\,\mathsf{G}^{(\mathrm{R})}(t,\alpha-\alpha')
	\label{ttaPrim}
\end{align}
Here we have used the fact that the $s$-wave modes $\upphi_0$ of the field $\upphi$ are projected out by integration over the circle
\be 
\upphi_0(t,r) = \frac{1}{2\pi}\intf{0}{2\pi}\mathrm{d}\alpha\,\upphi(t,r,\alpha)
\ee
and used the circular symmetry of the state $|\,\Omega\,\ra$ to write $\mathsf{G}^{(\mathrm{R})}_{(\text{$s$-wave})}(t)$ in terms of the full 3D one-sided retarded two-point function 
\be 
\mathsf{G}^{(\mathrm{R})}(t,\alpha) \equiv \mathrm{i}\,\Theta(t)\,\la\,\Omega\,\big|\,[\upphi(+t/2,r_{\mathrm{sh}},\alpha),\upphi(-t/2,r_{\mathrm{sh}},0)]\,\big|\,\Omega\,\big\ra
\label{G3D}
\ee
of the full 3D scalar field $\upphi$. \eqref{ttaPrim} can be further simplified to read
\begin{align}
	\mathsf{G}^{(\mathrm{R})}_{(\text{$s$-wave})}(t)
	&= \frac{1}{4\pi^2}\intf{0}{2\pi}\mathrm{d}\alpha\,\mathrm{d}\alpha'\,\mathsf{G}^{(\mathrm{R})}(t,\alpha-\alpha')\\
	&= \frac{1}{2\pi^2}\intf{0}{2\pi}\mathrm{d}\upalpha\intf{-\frac{\upalpha}{2}}{+\frac{\upalpha}{2}}\mathrm{d}\bar{\alpha}\,\mathsf{G}^{(\mathrm{R})}(t,\upalpha)\\
	&= \frac{1}{2\pi^2}\intf{0}{2\pi}\upalpha\,\mathrm{d}\upalpha\,\mathsf{G}^{(\mathrm{R})}(t,\upalpha)
	\label{tta}
\end{align}
Here we have defined 
\be 
\upalpha \equiv \alpha - \alpha', \qquad \bar{\alpha} \equiv \frac{\alpha + \alpha'}{2}
\ee 
and used the $\Z_2$ symmetry of the state $|\,\Omega\,\ra$ under $\upalpha \to -\upalpha$.

\begin{figure}[H]
	\begin{center}
		\includegraphics[scale=0.6]{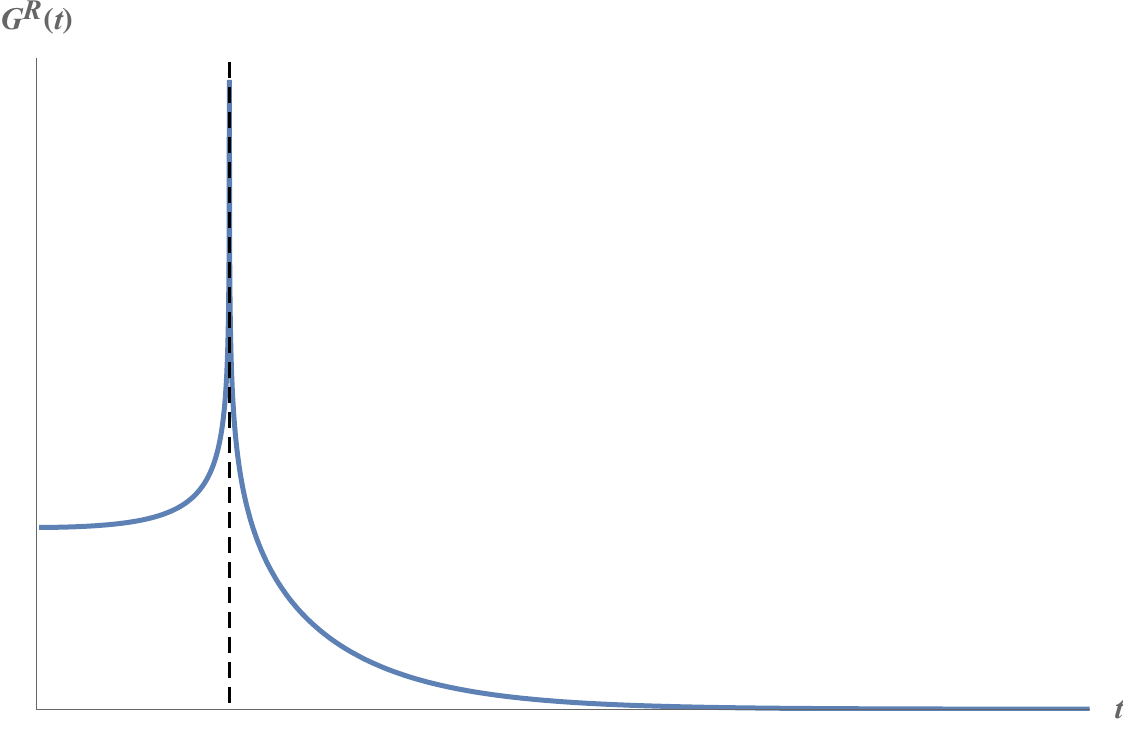}
		\caption{Numerical plot of \eqref{tta} with $m^2\ell^2 = 0.75$, $r_{\mathrm{sh}}=0.95\ell^2$ and in units in which $\ell = 1$.}
		\label{timelikem}
	\end{center}
\end{figure}

\begin{figure}[H]
	\begin{center}
		\includegraphics[scale=0.65]{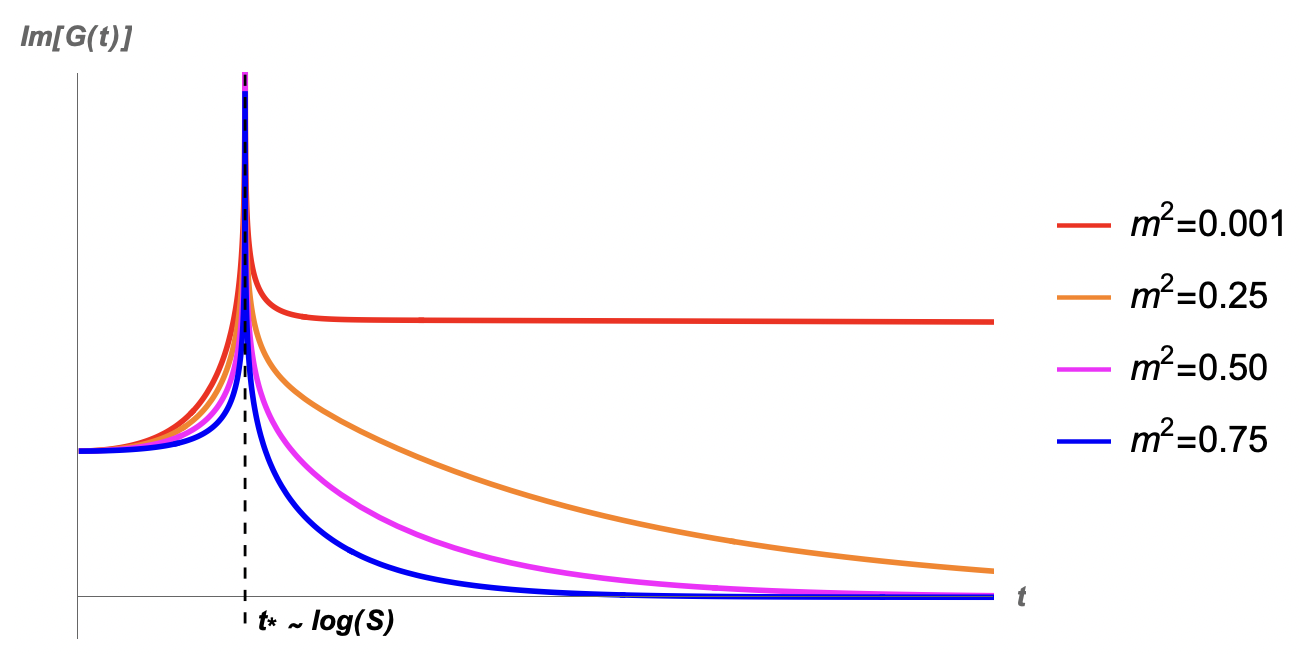}
		\caption{Numerical plot of \eqref{tta} for several choices of mass, again with $r_{\mathrm{sh}} = 0.95\ell^2$ and in units in which $\ell = 1$.}
		\label{multiple}
	\end{center}
\end{figure}

Since we are working at leading order in the semiclassical expansion, we can calculate \eqref{tta} by first calculating the Euclidean two-point function of $\upphi$ on the three-sphere and then analytically continuing the result back to Lorentz signature. For a massive scalar field of 3D mass $m$, the result is\footnote{More explicitly, we have \cite{Mirbabayi:2022gnl}
\begin{equation}
	\mathsf{G}^{(\mathsf{R})}(t,\alpha) = \Theta(\Delta\tau)\,\tilde{c}\cdot\tanh\inp{\frac{\Delta\tau(t,\alpha)}{2\ell}}\cdot {}_2F_1\inp{1-\Delta_+,1-\Delta_-;\frac{1}{2},-\sinh^2\inp{\frac{\Delta\tau(t,\alpha)}{2\ell}}}
\end{equation}
with 
\begin{equation}
	\tilde{c} = \frac{2\pi}{(4\pi)^{3/2}\,\Gamma\big(\frac{1}{2}\big)\ell^2}
\end{equation}
We will find it enough to work with the less explicit form \eqref{ttsolmassive} (in particular this is what we use for the numerics leading to figures \ref{timelikem} and \ref{multiple}.)
} (see e.g. \cite{Spradlin:2001pw} and references within)
\be
\mathsf{G}^{(\mathsf{R})}(t,\alpha) = -2\,\Theta(\Delta\tau)\,c(m^2)\cdot \mathrm{Im}\insb{{}_2F_1\inp{\Delta_+,\Delta_-;\frac{3}{2},\cosh^2\inp{\frac{\Delta\tau(t,\alpha)}{2\ell}}}}
\label{ttsolmassive}
\ee
where
\be
c(m^2) = \frac{\Gamma(\Delta_+)\Gamma(\Delta_-)}{(4\pi)^{3/2}\,\Gamma\big(\frac{3}{2}\big)\ell^2} \qquad\text{and}\qquad \Delta_{\pm} = 1 \pm \sqrt{1-m^2\ell^2}
\label{massiveparam}
\ee
and we have used an identity analogous to \eqref{RvsW}. In \eqref{ttsolmassive} above, $\Delta\tau(t,\alpha)$ denotes the proper time (signed geodesic distance) between the two insertion points in \eqref{tta}, which is given by 
\be 
\sinh^2\inp{\frac{\Delta\tau(t,\alpha)}{2\ell}} = f(r_{\mathrm{sh}})\,\sinh^2\inp{\frac{t}{2\ell}} -\inp{\frac{r_{\mathrm{sh}}}{\ell}}^2\sin^2\inp{\frac{\alpha}{2}}
\label{dtvstasym}
\ee
(see \S \ref{timelikeassym} for details). \eqref{ttsolmassive} has a mass-independent UV divergence 
\be 
\mathsf{G}^{(\mathsf{R})}(t,\alpha) \ \underset{\Delta\tau \to 0}{\sim} \ \frac{1}{2\pi}\frac{1}{\Delta\tau(t,\alpha)}
\label{lightcone}
\ee
on the lightcone $\Delta\tau(t,\alpha) = 0$; this is of course the usual UV contact/lightcone singularity of field theory. \eqref{ttsolmassive} is quoted for the case where $\Delta\tau(t,\alpha) \in \R$, corresponding to the case where the 3D events $(-t/2,r_{\mathrm{sh}},\alpha)$ and $(+t/2,r_{\mathrm{sh}},0)$ are causally separated. As usual, \eqref{ttsolmassive} vanishes when the 3D events $(-t/2,r_{\mathrm{sh}},\alpha)$ and $(+t/2,r_{\mathrm{sh}},0)$ are spacelike separated. \eqref{ttsolmassive} will hold for general $(t,\alpha)$ provided we define $\Theta(z)$ to vanish off of the positive real axis: $\Theta(z) = 0$ for $z \notin \R_+$.

We can now recover the expected exponential decay \eqref{G0(t)} using the deep IR (on cosmic scales) of our bulk. We begin with an analytic argument, which we will later corroborate with numerics below. Assuming that the late time limit $t \to \infty$ commutes with the homogenization \eqref{Ghomo}, we can use that
\begin{equation}
\mathrm{Im}\insb{{}_2F_1\inp{\Delta_+,\Delta_-;\frac{3}{2},\cosh^2\inp{\frac{\Delta\tau(t,\alpha)}{2\ell}}}} \ \underset{t\to\infty}{\sim} \ \exp\inp{-\frac{2\pi\Delta_-t}{2\pi\ell}}
\label{2F1decay}
\end{equation}
to find that, up to an overall coefficient which can be absorbed into the normalization of the field $\upphi$,
\begin{equation}
	G^{(\mathrm{R})}(t) \ \underset{t \to \infty}{\sim} \ e^{-2\pi\Delta_-\mathcal{T}t}
	\label{massive}
\end{equation}
where we have used \eqref{TdS} (as well as the equivalence of the bulk temperature and tomperature) to rewrite the exponents. This is the deep IR cosmic scale exponential decay of the retarded two-point function that was promised earlier.

The integral \eqref{tta} can be done numerically (for a fixed choice of $r_{\mathrm{sh}}$, e.g. $r_{\mathrm{sh}} = 0.95\,\ell$) and plotted. The result is given in figure \ref{timelikem} below for $m^2\ell^2 = 0.75$ and in figure \ref{multiple} below for several other choices of mass (to illustrate the generic nature of the main features).

It is interesting to study the correlator \eqref{tt}/\eqref{tta} as a function of (static patch) time $t$.  For any mass, we have the following universal features: There is a universal critical time 
\be 
t_{*} \equiv 2\,\ell\,\mathrm{arcsinh}\inp{\frac{r_{\mathrm{sh}}}{\ell\sqrt{f(r_{\mathrm{sh}})}}} \ \sim \ \beta_{\mathrm{GH}}\log(S)
\label{t*1}
\ee 
For $t < t_*$, the correlator grows with time, while for $t > t_*$, the correlator rapidly decays with time, consistent with the late time exponential decay expected based on the periodicity \eqref{period} and required in order to match with \eqref{G0(t)} and the analytic expression \eqref{massive} in the deep IR. The correlator diverges as one approaches the critical time $t = t_*$ from either side.

In order to understand the early time growth as well as the divergence as $t \to t_*$, it is helpful to first discuss the higher-dimensional role of the critical timescale $t_*$ in the context of the 3D proper time function $\Delta\tau(t,\alpha)$ defined by \eqref{dtvstasym} above. Up until the critical time \eqref{t*1}, there are 3D angles for which the 3D separation between the events $(-t/2,r_{\mathrm{sh}},\alpha)$ and $(+t/2,r_{\mathrm{sh}},0)$ is null or spacelike. Roughly, this occurs because the blueshift factor $f(r_{\mathrm{sh}})$ at the stretched horizon supresses timelike displacements $\mathrm{d}t$ relative to angular displacements $\mathrm{d}\alpha$ in the 3D proper time element 
\be 
\mathrm{d}\tau_{(\mathrm{3D})} = \sqrt{f(r_{\mathrm{sh}})\,\mathrm{d}t^2 - r_{\mathrm{sh}}^2\,\mathrm{d}\alpha^2}
\ee 
More specifically, for $t \leq t_*$, there is a maximal angular displacement\footnote{Note that while the angle $\alpha$ parameterizes a circle and therefore runs from $0$ to $2\pi$ (with $\alpha \sim \alpha + 2\pi$), the angular \textit{displacement} $\Delta \alpha$ only runs from $0$ to $\pi$ (inclusive) since any two points on the circle lying at angles $\alpha$ and $\alpha + \delta\alpha$ with $\delta \alpha > \pi$ are actually only ever $\Delta\alpha = 2\pi - \delta\alpha < \pi$ apart in angular displacement.} $\Delta\alpha_*(t)$, given by
\be 
\Delta\alpha_*(t) = 2\,\mathrm{arcsin}\inp{\frac{\sqrt{\ell f(r_{\mathrm{sh}})}}{r_{\mathrm{sh}}}\,\sinh\inp{\frac{t}{2\ell}}}
\label{amax}
\ee
beyond which the separation becomes spacelike.

Begin with the case $t < t_*$ so that $\Delta\alpha_*(t) < \pi$. Then, for $\alpha \text{ (mod $\pi$)} = \Delta\alpha_*(t)$, the 3D separation between the events $(-t/2,r_{\mathrm{sh}},\alpha)$ and $(+t/2,r_{\mathrm{sh}},0)$ is null, while for $\alpha \text{ (mod $\pi$)} > \Delta\alpha_*(t)$, the 3D separation between the events $(-t/2,r_{\mathrm{sh}},\alpha)$ and $(+t/2,r_{\mathrm{sh}},0)$ is spacelike. At the critical time $t = t_*$ the maximal angular displacement $\Delta\alpha_*(t_*) = \pi$ so the events $(-t/2,r_{\mathrm{sh}},\alpha)$ and $(+t/2,r_{\mathrm{sh}},0)$ are timelike separated for all angles except $\alpha = \pi$, for which the 3D separation is null. After the critical time $t > t_*$ the events $(-t/2,r_{\mathrm{sh}},\alpha)$ and $(+t/2,r_{\mathrm{sh}},0)$ are timelike separated for all angles.

We can now understand the 3D Green's function \eqref{G3D} as a function of $t$ and $\alpha$. For fixed $t \leq t_*$, the 3D retarded Green's function \eqref{tta} is supported for $\alpha < \Delta\alpha_*(t)$ and $\alpha > 2\pi-\Delta\alpha_*(t)$%, diverging at $\alpha = \Delta\alpha_*(t)$ and at $\alpha = 2\pi-\Delta\alpha_*(t)$ due to the lightcone singularity \eqref{lightcone}
. Indeed, for $t \leq t_*$ the integral \eqref{tta} truncates to 
\begin{align}
	G^{(\mathrm{R})}\big(t \leq t_*\big) 
	&= \frac{1}{2\pi^2}\inp{\int_0^{\Delta\alpha_*(t)}\alpha\,\mathrm{d}\alpha + \ \ \int_{2\pi-\Delta\alpha_*(t)}^{2\pi}\alpha\,\mathrm{d}\alpha}\mathsf{G}^{(\mathrm{R})}(t,\alpha)\\
	&= \frac{1}{\pi}\int_0^{\Delta\alpha_*(t)}\mathrm{d}\alpha\,\mathsf{G}^{(\mathrm{R})}(t,\alpha)
	\label{GhomoTr}
\end{align}
where we have used that 
\begin{align}
	\int_{2\pi-\Delta\alpha_*(t)}^{2\pi}\alpha\,\mathrm{d}\alpha\,\mathsf{G}^{(\mathrm{R})}(t,\alpha)
	&= \int_{0}^{\Delta\alpha_*(t)}\inp{2\pi-\alpha}\mathrm{d}\alpha\,\mathsf{G}^{(\mathrm{R})}(t,2\pi-\alpha)\\
	&= \int_{0}^{\Delta\alpha_*(t)}\inp{2\pi-\alpha}\mathrm{d}\alpha\,\mathsf{G}^{(\mathrm{R})}(t,\alpha)\\
	&= \int_{0}^{\Delta\alpha_*(t)}\inp{2\pi-\alpha}\mathrm{d}\alpha\,\mathsf{G}^{(\mathrm{R})}(t,\alpha)\\
	&= 2\pi\int_{0}^{\Delta\alpha_*(t)}\mathrm{d}\alpha\,\mathsf{G}^{(\mathrm{R})}(t,\alpha)-\int_{0}^{\Delta\alpha_*(t)}\alpha\,\mathrm{d}\alpha\,\mathsf{G}^{(\mathrm{R})}(t,\alpha)
\end{align}
which follows from the $\Z_2$ symmetry of the state $|\,\Omega\,\ra$ under $\alpha \to - \alpha$ and the $2\pi$ periodicity of the circle. So for early times, the 2D two-point function \eqref{tt} is simply given (up to normalization) by the angular homogenization of the 3D retarded two-point function \eqref{G3D}:
\be 
G^{(\mathrm{R})}(t\leq t_*) = \frac{1}{\pi}\intf{0}{\Delta\alpha_*(t)}\mathrm{d}\alpha\,\mathsf{G}^{(\mathrm{R})}(t,\alpha)
\label{Ghomo}
\ee

This allows us to understand the early time growth of $G^{(\mathrm{R})}(t)$: as $t$ gets closer and closer to $t_*$, the retarded correlation function \cref{GhomoTr} sees more and more of the internal circle---as well as more and more of the static patch (see section \ref{emergence} below)---and hence acesses more and more field modes and correlations. 

To see that \eqref{GhomoTr} diverges as $t \to t_*$, we can note the following\footnote{We thank D. Stanford for helpful discussions on this point.}: For fixed $t < t_*$, the lightcone singularity \eqref{lightcone} of the 3D correlator---and hence of the integrand of \eqref{GhomoTr}---takes the form
\be 
\mathsf{G}^{(\mathrm{R})}(t,\alpha) \ \underset{\alpha \to \Delta\alpha_*}{\sim} \ %\frac{\mathrm{i}}{4\pi\ell^2}\frac{\sqrt{2}}{r_{\mathrm{sh}}}\frac{1}{\sqrt{\sin\big(\Delta\alpha_*(t)\big)}}\,\frac{1}{\sqrt{\Delta\alpha_*(t)-\alpha}}
\frac{1}{\sqrt{\sin\big(\Delta\alpha_*(t)\big)}}\,\frac{1}{\sqrt{\Delta\alpha_*(t)-\alpha}}
\label{tlsing}
\ee  
This singularity is integrable since $\Delta\alpha_*(t) < \pi$ for $t < t_*$, explaining why the 2D correlator \eqref{GhomoTr} is finite for $t < t_*$. However, right at $t = t_*$, the nature of the singularity changes (as suggested by the factor of $1/\sqrt{\sin\big(\Delta\alpha_*(t)\big)}$ in \eqref{tlsing}): we find a simple pole 
\be 
\mathsf{G}^{(\mathrm{R})}(t_*,\alpha) \ \underset{\alpha \to \pi}{\sim} \ \ \frac{1}{\pi-\alpha}
\ee  
which leads to a divergence of the 2D correlator \eqref{GhomoTr} right at $t = t_*$. As a function of $t \to t_*$, we have
\be 
G^{(\mathrm{R})}(t) \ \underset{t \to t_*}{\sim} \ \frac{1}{\sqrt{|t_*-t|}}
\label{divt}
\ee

From the perspective of the bulk, we can understand this singularity as follows: We can interpret the 2D Green's function \eqref{t0} in terms of the emission and reabsorption amplitude of scalar Hawking quanta. The timescale \eqref{t*1} is indeed the timescale expected for a bulk Hawking quantum to be emitted from and reabsorbed by the horizon. In our specific context, the quanta originate from 3D scalar fields with zero angular momentum. The time \eqref{t*1} is precisely the time at which this emission-reabsorption can occur within the $s$-wave sector; In this sense it appears that the divergence \eqref{divt} corresponds to the on-shell propagation of a particle from the stretched horizon to the pode and back\footnote{We thank H. Lin for helpful discussions on this point.}. 

\subsection{A Comment on Bulk Emergence}
\label{emergence}
\quad \ The correspondence between \eqref{G0(t)} and \eqref{tt} provides an example of bulk emergence: The quantity \eqref{G0(t)} (or, rather, its analog for the possible dual of the bulk field \eqref{DCscalar}) is defined purely on the boundary in terms of boundary variables but is computed in the bulk in a way which probes increasingly deeply into the static patch. As explained in appendix \ref{AppGeodesics}, up until a time 
\be 
t_{*} \equiv 2\,\ell\,\mathrm{arcsinh}\inp{\frac{r_{\mathrm{sh}}}{\ell\sqrt{f(r_{\mathrm{sh}})}}} \ \sim \ \beta_{\mathrm{GH}}\log(S)
\label{tscr}
\ee 
(which would have been the scrambling time in an ordinary fast-scrambling theory at inverse temperature $\beta_{\mathrm{GH}}$ \cite{Hayden:2007cs,Sekino:2008he,Maldacena:2015waa}) the region of the bulk static patch which contributes to the correlator \eqref{G0(t)}/\eqref{tt} lies between the stretched horizon and the minimum radius
\be 
r_{\mathrm{min}}(t) \equiv  \frac{r_{\mathrm{sh}}\,\mathrm{sech}\big(\frac{t}{2\ell}\big)}{\sqrt{1-\inp{\frac{r_{\mathrm{sh}}}{\ell}}^2\tanh^2\big(\frac{t}{2\ell}\big) + f(r_{\mathrm{sh}})\tan^2\inp{\frac{\Delta\alpha_{*}(t)}{2}}}}
\label{rmin}
\ee
where $\Delta\alpha_{*}(t)$ is as defined as in \eqref{amax} above.

Past the critical time $t_*$, the 2D bulk correlator \eqref{tt} probes the entire bulk static patch. We can therefore think of the timescale \eqref{tscr}, in the context of this particular example of bulk emergence, as the critical timescale at which the full bulk spacetime emerges from the corresponding boundary degrees of freedom.

\begin{figure}[H]
	\begin{center}
		\scalebox{0.75}{%% Creator: Inkscape 1.0.1 (3bc2e813f5, 2020-09-07), www.inkscape.org
%% PDF/EPS/PS + LaTeX output extension by Johan Engelen, 2010
%% Accompanies image file 'cor.pdf' (pdf, eps, ps)
%%
%% To include the image in your LaTeX document, write
%%   \input{<filename>.pdf_tex}
%%  instead of
%%   \includegraphics{<filename>.pdf}
%% To scale the image, write
%%   \def\svgwidth{<desired width>}
%%   \input{<filename>.pdf_tex}
%%  instead of
%%   \includegraphics[width=<desired width>]{<filename>.pdf}
%%
%% Images with a different path to the parent latex file can
%% be accessed with the `import' package (which may need to be
%% installed) using
%%   \usepackage{import}
%% in the preamble, and then including the image with
%%   \import{<path to file>}{<filename>.pdf_tex}
%% Alternatively, one can specify
%%   \graphicspath{{<path to file>/}}
%% 
%% For more information, please see info/svg-inkscape on CTAN:
%%   http://tug.ctan.org/tex-archive/info/svg-inkscape
%%
\begingroup%
  \makeatletter%
  \providecommand\color[2][]{%
    \errmessage{(Inkscape) Color is used for the text in Inkscape, but the package 'color.sty' is not loaded}%
    \renewcommand\color[2][]{}%
  }%
  \providecommand\transparent[1]{%
    \errmessage{(Inkscape) Transparency is used (non-zero) for the text in Inkscape, but the package 'transparent.sty' is not loaded}%
    \renewcommand\transparent[1]{}%
  }%
  \providecommand\rotatebox[2]{#2}%
  \newcommand*\fsize{\dimexpr\f@size pt\relax}%
  \newcommand*\lineheight[1]{\fontsize{\fsize}{#1\fsize}\selectfont}%
  \ifx\svgwidth\undefined%
    \setlength{\unitlength}{293.12464225bp}%
    \ifx\svgscale\undefined%
      \relax%
    \else%
      \setlength{\unitlength}{\unitlength * \real{\svgscale}}%
    \fi%
  \else%
    \setlength{\unitlength}{\svgwidth}%
  \fi%
  \global\let\svgwidth\undefined%
  \global\let\svgscale\undefined%
  \makeatother%
  \begin{picture}(1,0.97243517)%
    \lineheight{1}%
    \setlength\tabcolsep{0pt}%
    \put(0,0){\includegraphics[width=\unitlength,page=1]{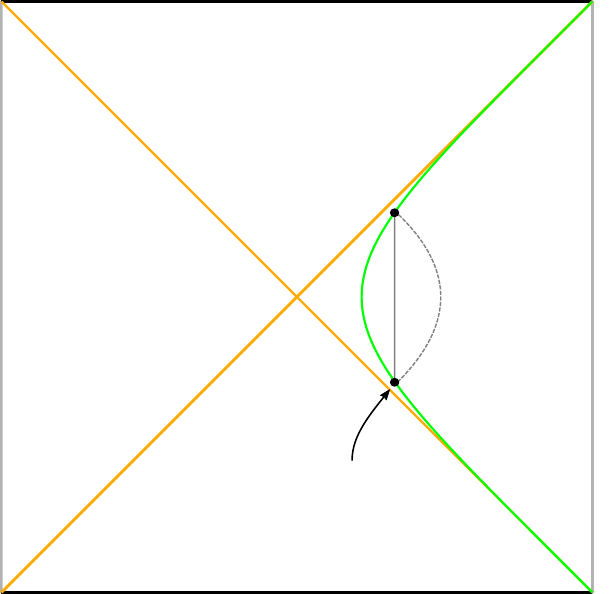}}%
    \put(0.50123498,0.17676189){\makebox(0,0)[lt]{\lineheight{1.25}\smash{\begin{tabular}[t]{l}$\phi(-t/2, r_{\mathrm{sh}})$\end{tabular}}}}%
    \put(0,0){\includegraphics[width=\unitlength,page=2]{cor.pdf}}%
    \put(0.49393872,0.77815913){\makebox(0,0)[lt]{\lineheight{1.25}\smash{\begin{tabular}[t]{l}$\phi(+t/2, r_{\mathrm{sh}})$\end{tabular}}}}%
    \put(0,0){\includegraphics[width=\unitlength,page=3]{cor.pdf}}%
  \end{picture}%
\endgroup%
}
		\caption{The correlation function \eqref{tt} is calculated using bundles of curves (i.e. the curves lying within the region shaded in grey) connecting the two insertion points (black dots). These curves descend from 3D geodesics with angular displacement between zero and $\Delta\alpha_*(t)$. The bundle reaches out to a maximal radius $r_{\mathrm{min}}(t)$ defined by \eqref{rmin}.}
		\label{cor}
	\end{center}
\end{figure}

\subsection{Implications for the Boundary-to-Bulk Mapping}
\label{bb}
\quad \ It is important to note the following mis-match between \eqref{G0(t)} and the 2D bulk correlator \eqref{tt}: The early time features of \eqref{tt}---namely the growth for $t < t_*$ and the divergence as $t \to t_*$---are completely absent from the boundary correlator \eqref{G0(t)}. These early time features encode propagation into the bulk, and help distinguish propagating bulk fields from trapped near-horizon degrees of freedom. Of course both propagating and near-horizon degrees of freedom will exhibit the late time thermal/quasinormal decay \eqref{G0(t)}. A related fact is that, according to \cite{Du:2004jt}, there are no Fermionic quasinormal modes in the $s$-wave sector in dS$_3$. This provides additional bulk evidence for the absence of propagating bulk fields corresponding to the fundamental Fermions $\chi_i$.

Susskind and Witten have argued \cite{Lenny} that this implies that the fundamental Fermions $\chi_i$ of the $\mathrm{DSSYK}_{\infty}$ model which appear in the boundary correlation function \eqref{G0(t)} overwhelmingly constitute near-horizon degrees of freedom\footnote{We thank L. Susskind and E. Witten for explaining this point to us.}. This means that, rather than having $\sim O(N)$ light bulk fields, we instead have a large number of many near-horizon degrees of freedom which contribute to the entropy but do not propogate, and perhaps a small number of bulk fields formed as collective excitations of the fundamental fermions. See section 9 of \cite{Lenny} for a more detailed discussion.

\section{Future Directions}
\label{discussion}

\subsection{The Charged DSSYK Model}
\label{U(1)}
\quad \ Throughout this paper, we have given a rather privileged position to the $(2+1)$-dimensional origin of the solution \eqref{metric},\eqref{fi=r},\eqref{dS2'} and action \eqref{action}. One may therefore ask: if this 3D origin is truly fundamental, is there a refinement of our model that allows us to capture more of this 3D parent theory?

A possible starting point is the following: In going from the 3D parent theory \eqref{3DCC} to the 2D theory \eqref{action} via the dimensional reduction \eqref{dimredansatz} (and similarly in going from the 3D matter action \eqref{3Dscalar} to the dilaton-coupled matter action \eqref{DCscalar}), we restricted ourselves to the $s$-wave sector of the 3D theory. We can include more information about the 3D parent spacetime by allowing our 3D fields to vary along the spatial circle on which we will eventually compactify. We can resolve such fields into Fourier modes via e.g. 
\be 
\upphi(x,\alpha) = \sum_{n\in\Z}\,\phi_{n}(x)\,e^{\mathrm{i}n\alpha}
\label{Fourier}
\ee
and as usual this will yield upon dimensional reduction infinite towers $\{\phi_n(\alpha)\}$ of 2D fields living within the dimensionally reduced $(1+1)$-dimensional theory. The circular spacetime symmetry $\alpha \to \alpha + \delta\alpha$ of the 3D parent theory will descend to a $U(1)$ internal symmetry 
\be 
\phi_n(x) \to e^{\mathrm{i}n\delta\alpha}\,\phi_n(x)
\label{U(1)sym}
\ee 
of the effective $(1+1)$-dimensional fields.

One can ask: is there a corresponding modification that one can make to the $\mathrm{DSSYK}_{\infty}$ model? One possibility is to consider the \textit{charged} $\mathrm{DSSYK}_{\infty}$ model. The charged $\mathrm{DSSYK}_{\infty}$ model, based on the charged SYK model introduced in \cite{Davison:2016ngz}, would describe \emph{Dirac} fermions $\upchi_i$ governed by the Hamiltonian 
\be 
H_{\mathrm{DS},\,\mathrm{charged}}[\upchi] \quad = \sum_{\substack{1\leq i_1<i_2\dots<i_{q/2}\leq N\\1\leq i_{q/2+1}<i_{q/2+2}\dots<i_q\leq N}}J_{i_1i_2\dots i_q}\,\adj{\upchi}_{i_1}\adj{\upchi}_{i_2}\dots\adj{\upchi}_{i_{q/2}}\upchi_{i_{q/2+1}}\upchi_{i_{q/2+2}}\dots\upchi_{i_q}
\label{charged}
\ee
with $q$ an even integer and with \emph{complex} couplings $J_{i_1i_2\dots i_q}$ obeying 
\be 
J_{i_1i_2\dots i_{q/2}i_{q/2+1}\dots i_q} = J_{i_{q/2+1}\dots i_qi_1i_2\dots i_{q/2}}^*
\label{complex}
\ee
drawn independently and at random from a Gaussian distribution of mean zero and of variance 
\be 
\big\la\,\|J_{i_1i_2\dots i_q}\|^2\,\big\ra_{\mathrm{c}} = \frac{1}{2}\frac{\big[(q/2)!\big]^2}{N^{q-1}}\,\mathcal{J}^2
\label{mcJcomplex}
\ee 
This model has a global $U(1)$ symmetry which may be capture the higher $(2+1)$-dimensional angular momentum modes \eqref{Fourier}. 

At least for the ordinary SYK model at low temperatures, it is known that adding charge instead gives rise to bulk electromagnetism \cite{Susskind:2020fiu}. If this were also the case at high temperatures, we would not be able to use this strategy to access higher angular momenta, but could instead use this strategy to include bulk propagating fields (i.e. the $U(1)$ gauge field) such as those studied in \eqref{2ptonesided}.

We leave an exploration of this possible extention of our conjectured duality for future work.

\subsection{Spacelike Observables} 
\label{spacelike}
\ \quad A key feature of de Sitter space is its exponential expansion. In fact, it is precisely this accelerated expansion which gives rise to the existence of the cosmological horizon which bounds and defines the static patch in the first place. However, the existence of this horizon also hides the exponentially growing region from any single observer/timelike worldline: In order to probe this behind-the-horizon region, we must use nonlocal observables with access to at least two points spacelike separated from one another by a horizon radius or more in proper distance.

A simple example of such an observable would be a spacelike ``behind-the-horizon" geodesic connecting the pode and antipode stretched horizons at equal height (value of the embedding space time $T$). A more exotic such observable would be a ``dilaton weighted geodesic" $\gamma$, which is an extremum of the dilaton-weighted length functional 
\be 
l_{\mathrm{dw}}[\gamma] = \int_{\gamma}\Phi\,\mathrm{d}s
\label{dw}
\ee
Following the logic of \cite{Susskind:2021esx} and \cite{Brown:2018bms}, we would expect the dilaton-weighted lengths to encode holographic complexity\footnote{From a $(2+1)$-dimensional perspective, this simply follows from the fact that dilaton-weighted geodesics are the dimensionally reduced avatars of maximal volume slices. From a $(1+1)$-dimensional perspective, this follows from the expectation that complexity should grow in a way which is sensitive to the number of degrees of freedom in the system \cite{Brown:2018bms}, which, for us, is encoded in the dilaton $\Phi$ (recall section \ref{entropy}); from this perspective, mapping complexity to the dilaton-weighted lengths of dilaton-weighted geodesics is just a generalization of the prescription given in \cite{Brown:2018bms} %for the wormhole solution of JT gravity with negative cosmological constant 
to the regime of positive cosmological constant and rapidly varying dilaton.} in our theory.

\begin{figure}[H]
	\begin{center}
		\includegraphics[scale=.35]{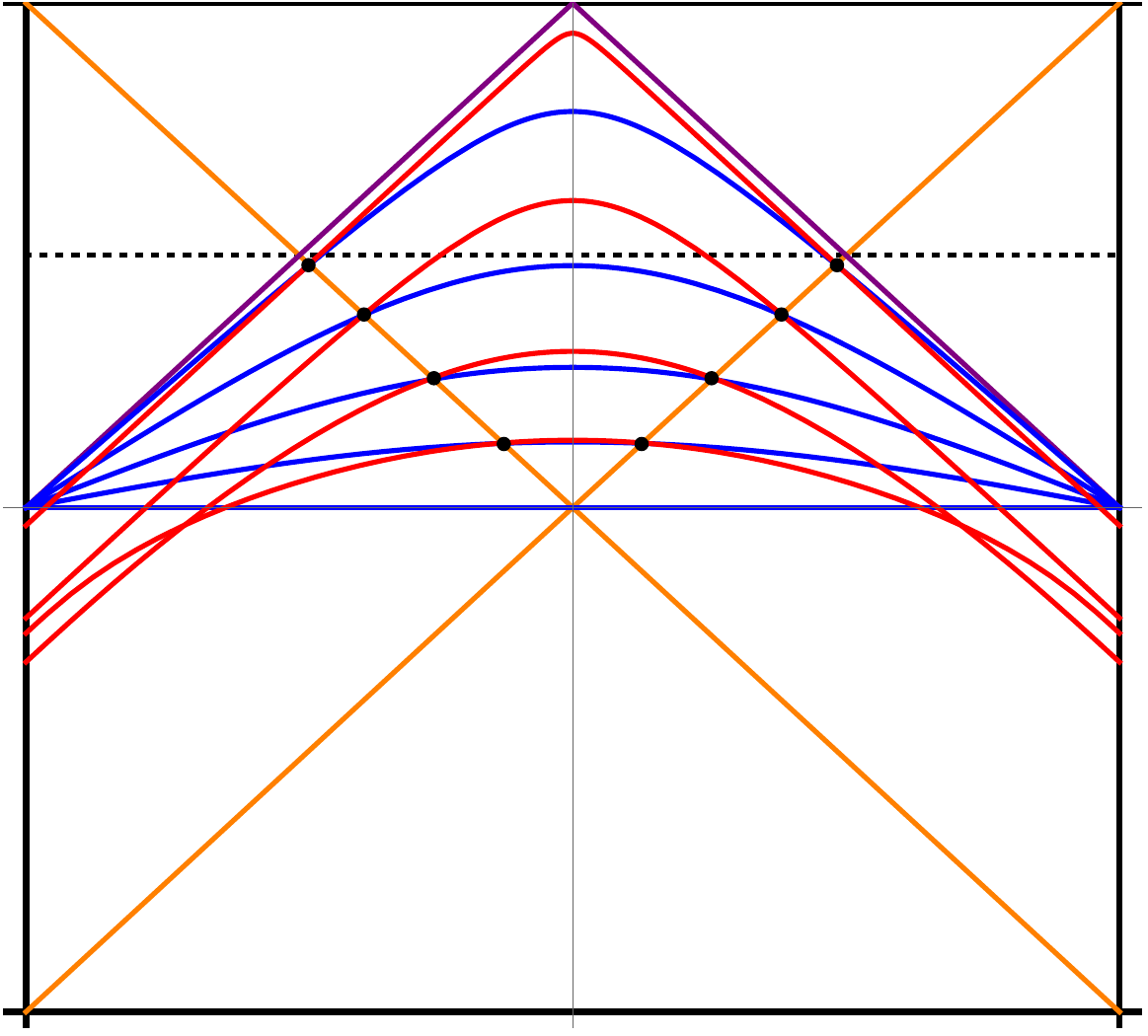}
		\caption{Dilaton-weighted geodesics (red) lie above ordinary geodesics (blue) in a way that becomes more and more pronounced as we increase $T$. However, as we approach the critical height $T \to T_* = r_{\mathrm{sh}}$ (dotted black line) they both approach the same pair of null segments joined at $\mathcal{I}^+$ (purple), albiet at different rates.}
		\label{geovsdilgeo}
	\end{center}
\end{figure}

Using the results of \cite{Jorstad:2022mls} one finds that both of these objects exhibit critical behavior at the same critical height $T_* = r_{\mathrm{sh}}$: the ordinary geodesics cease to exist and the dilaton-weighted lengths diverge%\footnote{This gives rise to another puzzle: A system of finite entropy $S$---such as de Sitter space in the context of static patch horizon holography---has a maximum possible complexity
. In terms of static patch time %\footnote{Here we have used the relation 
	%\be 
	%T(t) = \ell\,\sqrt{f(r_{\mathrm{sh}})}\,\sinh(t/\ell)
	%\label{T}
	%\ee 
	%between static patch time $t$ and embedding space height $T(t)$ of the stretched horizon.} 
$t$, the critical behavior for these spacelike observables appears at the exact same (up to a factor of $1/2$) timescale 
\be 
t_{*}^{(\mathrm{spacelike})} \equiv \ell\,\mathrm{arcsinh}\inp{\frac{r_{\mathrm{sh}}}{\ell\sqrt{f(r_{\mathrm{sh}})}}} \ \sim \ \beta_{\mathrm{GH}}\log(S)
\label{tscrdisc}
\ee 
as that found for bulk emergence (i.e. \eqref{tscr}) in section \ref{emergence}. The rates of growth of both the ordinary and dilaton-weighted lengths diverge at the critical time \eqref{tscrdisc} \cite{Jorstad:2022mls}. \cite{Susskind:2021esx} called this complexity growth ``hyperfast"; a key test of whether or not our conjectured duality accounts for behind-the-horizon bulk physics would be the existence (or lack thereof) of this same hyperfast complexity growth in the high temperature $\mathrm{DSSYK}_{\infty}$ model. We leave this for future work\footnote{In \cite{Susskind:2021esx} 
	it was suggested that hyperfast complexity growth might be implied by hyperfast scrambling. This is actually not always true. As a counterexample, we can consider a ``restricted" random quantum circuit model in which, at each time-step, the $N$ total qubits are split into $\frac{N}{q}$ groups of $q$, each of which is acted on by a random product of $q$ fermion operators $\psi_{i_1}\psi_{i_2}\dots\psi_{i_q}$% (viewed as an operator on the 1-qubit Hilbert space in the usual way)
	, with the various products chosen independently and at %(Haar-)
	random from the subspace of the unitary group consisting of products of $q$ fermion operators. At sufficiently high temperatures, we expect that this system satisfies all the criteria necessary to be well described by the epidemic model, and, therefore, to be a hyperfast scrambler at large $q$ (at least with the appropriately scaled Hamiltonian/clock). Let's now study its complexity growth. 
	
	A product of $q$ fermion operators has complexity $q$ \cite{qferm}. This is different from the complexity $4^q$ of an operator which is typical in all of $SU(2^q)$. The reason for this discrepancy is that a product of $q$ Fermions is a very specific type of operator in the zoo of $SU(2^q)$. The behavior of the complexity of this model, at high temperatures and at a time $O(1) < t \ll O(e^S)$, is thus expected to be
	\be
	\mathcal{C}_{\mathrm{restricted}} \ \sim \ \inp{\frac{N}{q}}\times q \times t = St
	\label{Crestricted}
	\ee 
	(where we have identified the high temperature entropy of the system as $S(\text{$N$ qubits}) = N$) leading to a late time linear growth with coefficient $S$. This could not have arisen from a period of hyperfast complexity growth at early times, and instead suggests ordinary fast complexity growth $\mathcal{C} \lesssim St$ at early times. We thank L. Susskind for discussions on this point.}.

One can also consider ``two-sided" correlation functions of bulk operators restricted to the pode and antipode horizons
\be
G(T) \equiv \la\,\Omega'\,|\,\phi_L(T)\,\phi_R(T)\,|\,\Omega'\,\ra
\label{LR}
\ee
where $\phi_L(T)$ and $\phi_R(T)$ denote operator insertions at the same height (value of $T$) along the stretched horizons of the antipode ($L$) and pode ($R$) static patches respectively---exactly the same as the boundary conditions for the geodesics and dilaton weighted geodesics described above. We leave an analysis of such correlation functions for future work, but end by mentioning that their ultimate behavior will provide another stress-test of whether or not our conjectured duality accounts for behind-the-horizon physics\footnote{We thank H. Lin for raising this possibility to us.}: Assume that the solution \eqref{metric},\eqref{fi=r},\eqref{dS2'} is really exactly described by the infinite temperature thermofield double state \eqref{TFD}. Then a left-right (antipode-pode) correlation function must be the same as a one-sided correlation function. We can see this by beginning with a finite (but possibly high) temperature thermofield double state, so that a left-right correlation function is simply an analytic continuation of a one-sided thermal correlation function
\begin{align}
	\la\,\mathcal{O}_R(t_R)\,\mathcal{O}_L(t_L)\,\ra_{\mathrm{TFD}_{\beta}} = \la\,\mathcal{O}(t_R-t_L + \mathrm{i}\beta/2)\,\mathcal{O}(0)\,\ra_{\beta}
	\label{acbeta}
\end{align}
where $\la\,\cdot\,\ra_{\mathrm{TFD}_{\beta}}$ denotes a (two-sided) correlation function in the thermofield double state at temperature $\beta$ and $\la\,\cdot\,\ra_{\beta}$ denotes a correlation function in the one-sided thermal state at temperature $\beta$. Taking the limit as $\beta \to 0$ gives
\begin{align}
	\la\,\mathcal{O}_R(t_R)\,\mathcal{O}_L(t_L)\,\ra_{\mathrm{TFD}_{0}} = \la\,\mathcal{O}(t_R-t_L)\,\mathcal{O}(0)\,\ra_{\beta = 0}
	\label{acbeta0}
\end{align}

If this does not hold in the bulk, it could be a sign that the solution \eqref{metric},\eqref{fi=r},\eqref{dS2'} is not dual to the pure infinite temperature thermofield double state, but rather a mixed state\footnote{One possibility is that the pode--antipode system is itself entangled with some other ``UV" degrees of freedom. These degrees of freedom might be ``UV" in the sense of not being visible in the semiclassical local description of \eqref{action} and/or may live in the Planck-sized region of the $t = 0$ slice lying between the two stretched horizons. We thank H. Lin for raising this possibility to us.} whose properties well approximate those of the infinite temperature TFD. A more pessimistic possibility would be that the infinite-temperature thermofield double state of $\mathrm{DSSYK}_{\infty}$ does indeed describe the pode--antipode system within the solution \eqref{metric},\eqref{fi=r},\eqref{dS2'}, but that it is only the static patches (and not the behind-the-horizon expanding region) which are holographically emergent. One nice fact which may be useful is that spacelike critical behavior appears when $t_R = -t_L = t_*^{(\mathrm{spacelike})}$, so that
\be 
t_R-t_L = 2\,t_*^{(\mathrm{spacelike})} = t_*^{(\mathrm{timelike})}
\label{svst}
\ee
Provided that the spacelike correlator \eqref{LR} exhibits the same critical behavior as the other spacelike observables considered above, \eqref{svst} allows for the consistency of the critical behaviors on the two sides of \eqref{acbeta0}.

A related issue is the following: one can define one-sided time evolution within the static patch (as we have done in section \ref{2ptonesided}) using the well-defined global boost generator $H_R-H_L$. The type of two-sided time evolution needed to define the spacelike observables mentioned in this section would require a ``global" Hamiltonian $H_R  + H_L$. In order to define this global Hamiltonian, we would require the additional input of a ``one-sided" Hamiltonian $H_L$ (or $H_R$). It is not clear whether or not the one-sided Hamiltonian is well-defined in the bulk in the semiclassical limit that we have been considering (see e.g. \cite{Chandrasekaran:2022cip} and \cite{Lenny}).

Understanding the issues and tests raised and discussed in this section will be extremely important for a better understanding of our conjectured duality. We leave this for future work.

\subsection{DSSYK for $\lambda > 0$ and the Two-Point Function}
\label{lgeq0}
\quad \ The two-point function \eqref{G(t)SYK} was calculated in the regime $\lambda \sim 0$. It will be of interest to understand how this calculation might generalize to finite\footnote{\cite{Berkooz:2018jqr} has studied $\mathrm{DSSYK}_{\infty}$ correlation functions at finite $\lambda$, but for operators of size $\sim O(q)$.} $\lambda$. It will also be important to understand the spectrum of the finite-$\lambda$ $\mathrm{DSSYK}_{\infty}$ model, and, in particular, if there are any collective modes corresponding to bulk fields. Assuming the theory does indeed possess bulk fields, it would be interesting to ask whether or not one can determine (e.g. numerically) the higher quasinormal modes of the two-point function of this hypothetical boundary collective mode since, in de Sitter space, the bulk quasinormal modes are integer spaced in de Sitter units \cite{Du:2004jt}, and it would be interesting to see if we can see this same physics in the boundary decription\footnote{We thank D. Stanford and Z. Yang for raising this point to us.}.\\

We leave all of this for future work.

\section*{Acknowledgements}

\quad \ I would like to thank Henry Lin, Raghu Mahajan, Douglas Stanford, Edward Witten, Zhenbin Yang, and Shunyu Yao for helpful discussions. I would especially like to thank Lenny Susskind for helpful discussions, encouragement to publish these results, help with some of the figures, feedback on the draft, and for coordinating the submission of our two related papers. I am supported in part by the Stanford Institute of Theoretical Physics, by NSF Grant PHY-1720397, by the NSF GRF Program under Grant No. DGE-1656518, and by a Fletcher Jones Foundation National Science Foundation Graduate Fellowship in the Stanford School of Humanities \& Sciences.

\appendix
\section{Timelike Geodesics in the Static Patch}
\label{AppGeodesics}
\quad \ In this appendix we provide some details regarding the calculation of the proper time \eqref{dtvstasym} quoted in the main text. We begin by studying, as a warm-up exercise, spherically symmetric geodesics connecting two times at fixed angle on the pode stretched horizon in dS$_D$. We then relax the assumption of spherical symmetry while simultaneously specializing to the case $D = 3$ in order to find the desired proper time \eqref{dtvstasym}.

\subsection{Spherically-Symmetric Geodesics}
\quad \ We begin by studying spherically-symmetric geodesics in dS$_D$ which are emitted from the (e.g.) pode stretched horizon at static patch time $-t_{\mathrm{e}}/2$ and recieved at the same (pode) stretched horizon at time $+t_{\mathrm{e}}/2$. As is well known (and can be easily checked) such geodesics are simply lines of constant global spatial polar coordinate $x$---related to static patch coordinates via \eqref{rglobal}
\be
\ell\cos\big(x(t,r)\big) = \frac{r}{\sqrt{1 + f(r)\sinh^2(t/\ell)}}
\label{xApp}
\ee
---which are affinely parameterized by the global time $\tau$, related to static patch coordinates via \eqref{tglobal}
\be 
\sinh^2\inp{\frac{\tau(t,r)}{\ell}} = f(r)\sinh^2\inp{\frac{t}{\ell}}
\label{tauvstApp}
\ee
where we remind the reader that 
\be 
f(r) = 1-\frac{r^2}{\ell^2}
\label{f(r)App}
\ee

The proper time $\Delta \tau$ between the events $(-t_{\mathrm{e}}/2,r_{\mathrm{sh}})$ and $(+t_{\mathrm{e}}/2,r_{\mathrm{sh}})$ is therefore easily determined to be twice the value of the global time at the event $(+t_{\mathrm{e}}/2,r_{\mathrm{sh}})$, i.e. 
\be 
\sinh^2\inp{\frac{\Delta\tau}{2\ell}} = f(r_{\mathrm{sh}})\sinh^2\inp{\frac{t_{\mathrm{e}}}{2\ell}}
\label{dtauvstApp}
\ee
Note for future reference that 
\be 
\sinh\inp{\frac{\Delta\tau}{2\ell}} \ \sim \ \frac{1}{2}\,\sqrt{f(r_{\mathrm{sh}})}\,\exp\inp{\frac{t_{\mathrm{e}}}{2\ell}}
\label{dtlate}
\ee 
as $t_{\mathrm{e}} \to \infty$.

We can use \eqref{xApp} to determine the profile of the geodesic in static patch coordinates as
\be
r(t) = \frac{r_0}{\sqrt{1 -f(r_0)\tanh^2(t/\ell)}}
\label{prof}
\ee
where we have defined $r_0 \equiv r(t = 0) = \ell\cos(x)$, which can be determined for a given $t_{\mathrm{e}}$ by plugging in the boundary condition $r(\pm t_{\mathrm{e}}/2) = r_{\mathrm{sh}}$. Doing this, we find that
\be 
r_0 = \frac{r_{\mathrm{sh}}\,\mathrm{sech}(t_{\mathrm{e}}/2\ell)}{\sqrt{1-\inp{\frac{r_{\mathrm{sh}}}{\ell}}^2\tanh^2(t_{\mathrm{e}}/2\ell)}}
\label{r0tau}
\ee

\subsection{Spherically Asymmetric 3D Geodesics}
\label{timelikeassym}
\quad \ We now relax the assumption of spherical symmetry while also specializing to the case $D = 3$ in which case %\eqref{static} 
the metric reduces to \eqref{dS3}. We now want to find the timelike geodesic in the static patch connecting the events $(-t_{\mathrm{e}}/2,r_{\mathrm{sh}},-\Delta\alpha/2)$ and $(+t_{\mathrm{e}}/2,r_{\mathrm{sh}},+\Delta\alpha/2)$ on the pode stretched horizon. We can represent this geodesic by a pair of functions $\big(r(t),\alpha(t)\big)$ which will together comprise an extremum of the proper time functional
\be 
\Delta\tau[r(t),\alpha(t)] = \int\mathrm{d}t\,\sqrt{f\big(r(t)\big)-\frac{\dot{r}(t)^2}{f\big(r(t)\big)}-r(t)^2\dot{\alpha}(t)^2}
\label{tauasymm}
\ee

The Lagrangian (integrand) does not explicitly depend on $t$ or $\alpha$ and so has a conserved energy\footnote{\eqref{Etasymm} can be calculated either using
	\be 
	E = -\dot{r}\,\pd{L}{\dot{r}} + L
	\label{EL}
	\ee 
	with $L(r)$ the Lagrangian (integrand) of \eqref{tauasymm} or using
	\be 
	E = -g_{ab}t^au^b
	\label{Et}
	\ee
	with $t^a = (\partial/\partial t)^a$ the timelike ``boost" Killing vector and $u^a$ the unit tangent to the geodesic.}
\be 
E = \frac{f(r)}{\sqrt{f(r)-\frac{\dot{r}^2}{f(r)}-r^2\dot{\alpha}^2}}
\label{Etasymm}
\ee 
%(which can be calculated using either method from the previous subsection) 
and a conserved angular momentum\footnote{\eqref{Jasymm} can be calculated either using 
	\be 
	J = \pd{L}{\dot{\alpha}}
	\label{JL}
	\ee 
	with $L(r)$ the Lagrangian (integrand) of \eqref{tauasymm} or using
	\be 
	J = +g_{ab}\,\alpha^au^b
	\label{Jalpha}
	\ee 
	with $\alpha^a = (\partial/\partial\alpha)^a$ the angular $U(1)$ ``rotation" Killing vector and $u^a$ the unit tangent to the geodesic.}
\be 
J = \frac{r^2\dot{\alpha}}{\sqrt{f(r)-\frac{\dot{r}^2}{f(r)}-r^2\dot{\alpha}^2}}
\label{Jasymm}
\ee 
We can use \eqref{Jasymm} to solve for $r^2\dot{\alpha}^2$ as
\be 
r^2\dot{\alpha}^2 = \inp{\frac{J^2}{r^2 + J^2}}\inp{f(r)-\frac{\dot{r}^2}{f(r)}}
\label{r2da2}
\ee
which we can then plug into \eqref{Etasymm} to give
\be 
\frac{E}{\sqrt{1+\frac{J^2}{r^2}}} = \frac{f(r)}{\sqrt{f(r)-\frac{\dot{r}^2}{f(r)}}}
\label{Er}
\ee

For $t \geq 0$ (so that $\dot{r} \geq 0$) \eqref{Er} can be rearranged to read 
\be 
\dot{r} = f(r)\sqrt{1 - \inp{1 + \frac{J^2}{r^2}}\frac{f(r)}{E^2}}
\label{dr}
\ee
%which reduces to \eqref{COEtau} when $J = 0$, as expected.
\eqref{dr} only depends on the angular displacement $\Delta\alpha$ implicitly through the angular momentum parameter $J$ so the problem of solving for the profile $r(t)$ is symmetric about $t = 0$. We must therefore have that $\dot{r}(0) = 0$, so we can evaluate \eqref{Er} at $t = 0$ to find
\be 
E = \sqrt{\inp{1 + \frac{J^2}{r_0^2}}f(r_0)}
\label{E0J}
\ee
\eqref{dr} can therefore be written in terms of the parameter $r_0$ as
\be 
\dot{r} = f(r)\sqrt{1 - \frac{\inp{1 + \frac{J^2}{r^2}}f(r)}{\inp{1 + \frac{J^2}{r_0^2}}f(r_0)}}
\label{drr0}
\ee 
We can throw out the case $\dot{r} = 0$ since this does not solve the actual equation of motion except in the edge case $r(t) = 0$ when $t_{\mathrm{e}} = \infty$ (which we will automatically recover below). \eqref{drr0} can therefore be separated and integrated to give 
%\be 
%t = \mathrm{arccoth}\inp{\sqrt{\frac{f(r_0)}{r_0^2 + J^2}}\sqrt{\frac{r_0^2\,r^2 + J^2}{f(r_0)-f(r)}}}
%\label{t(r)J}
%\ee
%where $t(r)$ is the inverse function of $r(t)$ restricted to the interval $r_0 \leq r \leq r_{\mathrm{sh}}$. We can then solve for the original profile $r(t)$ as
%or
\be 
r(t) = \sqrt{\frac{r_0^4 + J^2\inp{r_0^2 + f(r_0)\,^2\tanh^2(t/\ell)}}{r_0^2\inp{1-f(r_0)\tanh^2(t/\ell)} + J^2}}
\label{r(t)r0J}
\ee
which is symmetric about $t = 0$ and therefore defines $r(t)$ for all $-t_{\mathrm{e}}/2 \leq t \leq +t_{\mathrm{e}}/2$. Plugging in the boundary condition
\be 
r(\pm t_{\mathrm{e}}/2) = r_{\mathrm{sh}}
\label{bc}
\ee 
gives the condition
\begin{multline}
	r_0^4\Big[1-\inp{\frac{r_{\mathrm{sh}}}{\ell}}^2\tanh^2(t_{\mathrm{e}}/2\ell)\Big] -  r_0^2\Big[\inp{r_{\mathrm{sh}}^2-J^2}\mathrm{sech}^2(t_{\mathrm{e}}/2\ell)\Big]\\ + J^2\inp{\ell^2\tanh^2(t_{\mathrm{e}}/2\ell)-r_{\mathrm{sh}}^2} = 0
	\label{r04}
\end{multline}
Note that consistency of \eqref{r04} tells us that both $r_0 \to 0$ and $J \to 0$ as $t_{\mathrm{e}} \to \infty$. 

It will turn out to be more convenient to solve \eqref{r04} for $J$ in terms of $r_0$ and hyperbolic functions of $t_{\mathrm{e}}/2$ (instead of for $r_0$ in terms of $J$ and hyperbolic functions of $t_{\mathrm{e}}/2$). This gives 
\be 
J^2 = \frac{r_0^2\insb{r_{\mathrm{sh}}^2\,\mathrm{sech}^2(t_{\mathrm{e}}/2\ell)-r_0^2\inp{1-\inp{\frac{r_{\mathrm{sh}}}{\ell}}^2\tanh^2(t_{\mathrm{e}}/2\ell)}}}{\inp{\ell^2\tanh^2(t_{\mathrm{e}}/2\ell) - r_{\mathrm{sh}}^2 + r_0^2\,\mathrm{sech}^2(t_{\mathrm{e}}/2\ell)}}
\label{J2}
\ee
Note that, as expected, \eqref{r(t)r0J} reduces to \eqref{prof} and \eqref{r04}---or, equivalently, \eqref{J2}---yields \eqref{r0tau} when $J = 0$.

We can use \eqref{dr} to simplify \eqref{r2da2} as
\be 
\dot{\alpha} = \frac{J}{E}\frac{f(r)}{r^2}
\label{da}
\ee
The problem of solving for the profile $\alpha(t)$ is antisymmetric about $t = 0$, so we must have\footnote{We can integrate \eqref{da} directly using \eqref{r(t)r0J} and the boundary condition that $\alpha(-t_{\mathrm{e}}/2) = -\Delta\alpha/2$, which gives
	\be 
	\alpha(t)  = \arctan\inp{\frac{J}{r_0}\sqrt{\frac{\ell^2-r_0^2}{J^2 + r_0^2}}\,\,\tanh(t/\ell)}
	\label{a(t)r0J}
	\ee
	The would-be integration constant
	\be 
	\alpha_0 = \arctan\inp{\frac{J}{r_0}\sqrt{\frac{\ell^2-r_0^2}{J^2 + r_0^2}}\,\,\tanh(t_{\mathrm{e}}/2\ell)} - \frac{\Delta\alpha}{2}
	%\arctan\inp{\frac{J}{E}\frac{f(r_0)}{r_0^2}\,\tanh(t_{\mathrm{e}}/2)} - \frac{\Delta\alpha}{2}
	\label{a-}
	\ee
	can be shown to vanish after using the future boundary condition $\alpha(+t_{\mathrm{e}}/2) = +\Delta\alpha/2$, confirming that $\alpha(0) = 0$ as anticipated above. \eqref{a(t)r0J} then tells us that
	%\be 
	%\tan\inp{\frac{\Delta\alpha}{2}} = \frac{J}{E}\frac{f(r_0)}{r_0^2}\,\tanh(t_{\mathrm{e}}/2)
	%\label{dalphaE}
	%\ee
	%or, plugging in \eqref{E0J}, 
	\be 
	\tan\inp{\frac{\Delta\alpha}{2}} = \frac{J}{r_0}\sqrt{\frac{\ell^2-r_0^2}{J^2 + r_0^2}}\,\tanh(t_{\mathrm{e}}/2\ell)
	\label{dalpha}
	\ee
	Equality of \eqref{dalpha} and \eqref{dalphaindirect}---and, more generally, equality of \eqref{a(t)r0J} and \eqref{a(r)}---is guaranteed by \eqref{r(t)r0J} (along with the boundary condition \eqref{bc}).} that $\alpha(t = 0) = 0$. We can therefore integrate \eqref{da} via
\begin{align} 
	\alpha(t>0) 
	&= \int_{0}^{\alpha}\mathrm{d}\alpha \\
	&= \frac{J}{E}\int_{0}^{t}\mathrm{d}t\,\frac{f(r)}{r^2} \\
	&= \frac{J}{E}\int_{r_0}^{r}\frac{\mathrm{d}r}{\dot{r}}\,\frac{f(r)}{r^2} \\
	&= J\int_{r_0}^{r}\frac{\mathrm{d}r}{r^2\sqrt{\inp{1 + \frac{J^2}{r_0^2}}f(r_0) - \inp{1 + \frac{J^2}{r^2}}f(r)}} \\
	&= \arctan\inp{\frac{\ell J}{r_0}\sqrt{\frac{r^2-r_0^2}{r_0^2r^2 + \ell^2J^2}}}
	\label{a(r)}
\end{align}
This then tells us that 
\be 
\tan\inp{\frac{\Delta\alpha}{2}} = \frac{\ell J}{r_0}\sqrt{\frac{r_{\mathrm{sh}}^2-r_0^2}{\ell^2J^2 + r_0^2r_{\mathrm{sh}}^2 }}
\label{dalphaindirect}
\ee
In particular \eqref{dalphaindirect} tells us that the ratio
\be 
\frac{\ell J}{r_0^2} \ \sim \ \tan\inp{\frac{\Delta\alpha}{2}} %\times\inp{1 + O(r_0^2/r_{\mathrm{sh}}^2)}
\label{Jr02late}
\ee  
as $t_{\mathrm{e}} \to \infty$. Plugging this into \eqref{J2} then tells us that
\be 
r_0^2 \ \underset{t_{\mathrm{e}} \to \infty}{\sim} \ \frac{4\,r_{\mathrm{sh}}^2}{f(r_{\mathrm{sh}})}\,\cos^2\inp{\frac{\Delta\alpha}{2}}\,e^{-t_{\mathrm{e}}/\ell}
\label{r02late}
\ee

We can use \eqref{J2} to rewrite \eqref{dalphaindirect} as 
\be 
\tan^2\inp{\frac{\Delta\alpha}{2}} = \frac{\ell^2r_{\mathrm{sh}}^2\,\mathrm{sech}^2(t_{\mathrm{e}}/2\ell)-r_0^2\inp{\ell^2-r_{\mathrm{sh}}^2\tanh^2(t_{\mathrm{e}}/2\ell)}}{r_0^2\inp{\ell^2-r_{\mathrm{sh}}^2}}
\label{dar0}
\ee  
which gives 
\be 
r_0 = \frac{r_{\mathrm{sh}}\,\mathrm{sech}(t_{\mathrm{e}}/2\ell)}{\sqrt{1-\inp{\frac{r_{\mathrm{sh}}}{\ell}}^2\tanh^2(t_{\mathrm{e}}/2\ell) + \inp{1-\inp{\frac{r_{\mathrm{sh}}}{\ell}}^2}\tan^2(\frac{\Delta\alpha}{2})}}
\label{r0alpha}
\ee
This reduces to \eqref{r0tau} when $\Delta\alpha = 0$ and also yields \eqref{r02late} as $t_{\mathrm{e}} \to \infty$, as expected. 

We can then plug \eqref{r0alpha} into \eqref{J2} to find
\be 
\frac{J}{\ell} = \frac{r_{\mathrm{sh}}^2\sin(\Delta\alpha)}{\sqrt{\big(\inp{\ell^2-r_{\mathrm{sh}}^2}\cosh(t_{\mathrm{e}}) + r_{\mathrm{sh}}^2\cos(\Delta\alpha)\big)^2-\ell^4}}
\label{Jda}
\ee 
This vanishes when $\Delta\alpha = 0$ as expected. Most importantly, we can use \eqref{r0alpha} and \eqref{Jda} to confirm \eqref{r02late} and \eqref{Jr02late} respectively.

%We can finally plug both \eqref{r0alpha} and \eqref{Jda} into \eqref{r(t)r0J} to find that 
%\be 
%r(t) = 
%\label{r(t)asymm}
%\ee 
%\be 
%\alpha(t) = 
%\label{a(t)asymm}
%\ee 
%{\color{red}[TO DO: DO THIS]}

We can now calculate the proper time elapsed along the geodesic %in two ways. The first is by straightforwardly plugging \eqref{r(t)asymm}, \eqref{a(t)asymm} into \eqref{tauasymm} to find \eqref{dtauvstApp} {\color{red}[TO DO: DO THIS]}. The second is 
by using \eqref{r2da2} and \eqref{drr0} to find
\bea 
\Delta\tau 
\eq \int\mathrm{d}t\,\sqrt{f(r)-\frac{\dot{r}^2}{f(r)}-r^2\dot{\alpha}^2} \cr 
\eq 2\int_{r_0}^{r_{\mathrm{sh}}}\frac{\mathrm{d}r}{\dot{r}}\,\sqrt{\inp{\frac{r^2}{r^2 + J^2}}\inp{f(r)-\frac{\dot{r}^2}{f(r)}}} \cr
\eq 2\int_{r_0}^{r_{\mathrm{sh}}}\frac{\mathrm{d}r}{\sqrt{\inp{1 + \frac{J^2}{r_0^2}}f(r_0)-\inp{1 + \frac{J^2}{r^2}}f(r)}} \cr
\eq 
2\,\ell\,\mathrm{arcsinh}\inp{\frac{\sqrt{\inp{\frac{r_{\mathrm{sh}}}{r_0}}^2-1}}{\sqrt{1 + \frac{\ell^2J^2}{r_0^4}}}}
\label{dtasymm}
\eea
%which reduces to \eqref{dtsym} for $J = 0$, as expected. 
Plugging \eqref{r0alpha}, \eqref{Jda} into \eqref{dtasymm}, we find 
\be 
\boxed{\sinh^2\inp{\frac{\Delta\tau}{2\ell}} = f(r_{\mathrm{sh}})\,\sinh^2\inp{\frac{t_{\mathrm{e}}}{2\ell}} -\inp{\frac{r_{\mathrm{sh}}}{\ell}}^2\sin^2\inp{\frac{\Delta\alpha}{2}}}
\label{dtvstasymApp}
\ee
which reduces to \eqref{dtauvstApp} for $\Delta\alpha = 0$ as expected.

As $t_{\mathrm{e}} \to \infty$, we again find the behavior \eqref{dtlate}: All dependence on the angular displacement becomes (relatively) exponentially supressed at late times. We can see this more sugestively as follows: Using \eqref{Jr02late} we find that 
\be 
\sqrt{1 + \frac{\ell^2J^2}{r_0^4}} \ \sim \ \sec\inp{\frac{\Delta\alpha}{2}}
\label{den=sec}
\ee
and using \eqref{r02late} we find that
\be 
\sqrt{\inp{\frac{r_{\mathrm{sh}}}{r_0}}^2-1} \ \sim \ \frac{1}{2}\,\sqrt{f(r_{\mathrm{sh}})}\,\sec\inp{\frac{\Delta\alpha}{2}}\,\exp\inp{\frac{t_{\mathrm{e}}}{2\ell}}
\label{num=cos}
\ee
which makes it clear that the dependence on $\Delta\alpha$ cancels at leading order in \eqref{dtasymm}. 

For later use, we will want to calculate the late time limit of $\coth(\Delta\tau/\ell)$. Using that
\bea
\coth\inp{\frac{\Delta\tau}{\ell}} 
\eq \frac{2\sinh^2\frac{\Delta\tau}{2\ell} + 1}{2\sinh\frac{\Delta\tau}{2\ell}\sqrt{\sinh^2\frac{\Delta\tau}{2\ell} + 1}} %\cr
%\bn \cr
%\eq \frac{2\inp{\frac{r_{\mathrm{sh}}}{r_0}}^2 + \frac{J^2}{r_0^4} + 1}{2\sqrt{\inp{\frac{r_{\mathrm{sh}}}{r_0}}^2-1}\sqrt{\inp{\frac{r_{\mathrm{sh}}}{r_0}}^2 + \frac{J^2}{r_0^4}}}
\label{coth}
\eea
we find that 
\be 
\coth\inp{\frac{\Delta\tau}{\ell}} \ \sim \ 1+\frac{2}{{f(r_{\mathrm{sh}})}^2}\,\exp\inp{-\frac{4t_{\mathrm{e}}}{2\pi\ell}}
\label{coth(dt)lateApp}
\ee 
as $t_{\mathrm{e}} \to \infty$.

At early times
\be 
t_{\mathrm{e}} \ \leq \  t_{*} \equiv \ell\,\mathrm{arcsinh}\inp{\frac{r_{\mathrm{sh}}}{\ell\sqrt{f(r_{\mathrm{sh}})}}} \ \sim \ \ell\log(S)
\label{tscrApp}
\ee
demanding that our geodesic be timelike places a bound on the maximal angular displacement\footnote{Note that while the angle $\alpha$ parameterizes a circle and therefore runs from $0$ to $2\pi$ (with $\alpha \sim \alpha + 2\pi$), the angular \textit{displacement} $\Delta \alpha$ only runs from $0$ to $\pi$ since two points on the circle lying at angles $\alpha$ and $\alpha + \delta\alpha$ with $\delta \alpha > \pi$ are actually only ever $\Delta\alpha = 2\pi - \delta\alpha < \pi$ apart in angular displacement.} $\Delta\alpha_{\mathrm{m}}(t_{\mathrm{e}})$ possible for a given $t_{\mathrm{e}}$. The bound follows from the positivity of the proper time in \eqref{dtvstasymApp} and gives 
\be 
\sin^2\inp{\frac{\Delta\alpha}{2}} \ < \ \sin^2\inp{\frac{\Delta\alpha_{\mathrm{m}}(t_{\mathrm{e}})}{2}}
\label{damax}
\ee
where 
\be 
\sin^2\inp{\frac{\Delta\alpha_{\mathrm{m}}(t_{\mathrm{e}})}{2}} \equiv \frac{\ell^2f(r_{\mathrm{sh}})}{r_{\mathrm{sh}}^2}\,\sinh^2\inp{\frac{t_{\mathrm{e}}}{2\ell}}
\label{am}
\ee
Plugging \eqref{damax} into \eqref{r0alpha} we find that the smallest radius that can be probed at a given $t_{\mathrm{e}} \leq t_*$ is bounded below by
\be 
\boxed{r_{\mathrm{min}} \equiv  \frac{r_{\mathrm{sh}}\,\mathrm{sech}\big(\frac{t_{\mathrm{e}}}{2\ell}\big)}{\sqrt{1-\inp{\frac{r_{\mathrm{sh}}}{\ell}}^2\tanh^2\big(\frac{t_{\mathrm{e}}}{2\ell}\big) + f(r_{\mathrm{sh}})\tan^2\inp{\frac{\Delta\alpha_{\mathrm{m}}(t_{\mathrm{e}})}{2}}}}}
\label{r0min}
\ee

\bibliographystyle{jhep}
\bibliography{JT-DSSYK-V2} 

\end{document}